\documentclass[12pt]{spieman}  
\usepackage{amsmath,amsfonts,amssymb}
\usepackage{graphicx}
\usepackage{setspace}
\usepackage{tocloft}
\usepackage{pdflscape}
\usepackage{float}

\title{Measurement of telescope transmission using a Collimated Beam Projector}

\author[a,*]{Nicholas Mondrik}
\author[b]{Michael Coughlin}
\author[c]{Marc Betoule}
\author[c]{S\'{e}bastien Bongard}
\author[d]{Joseph P. Rice}
\author[d]{Ping-Shine Shaw}
\author[a,e]{Christopher W. Stubbs}
\author[d]{John T. Woodward}
\author[a,b,c,e]{LSST Dark Energy Science Collaboration}

\affil[a]{Department of Physics, Harvard University, 17 Oxford St., Cambridge, MA 02138, USA}
\affil[b]{School of Physics and Astronomy, University of Minnesota, Minneapolis, Minnesota 55455, USA}
\affil[c]{LPNHE, CNRS-IN2P3 and Universit\'{e} Paris 6, 4 place Jussieu, 75252 Paris Cedex 05, France}
\affil[d]{National Institute of Standards and Technology, 100 Bureau Dr., Gaithersburg, Maryland 20899, USA}
\affil[e]{Department of Astronomy, Harvard University, 60 Garden St., Cambridge, MA 02138, USA}

\newcommand{\sd}{StarDICE }

\cftpagenumbersoff{figure}
\cftpagenumbersoff{table}
\begin{document} 
\maketitle

\begin{abstract}
With the increasingly large number of type Ia supernova being detected by
current-generation survey telescopes, and even more expected with the upcoming
Rubin Observatory Legacy Survey of Space and Time, the precision of cosmological
measurements will become limited by systematic uncertainties in flux calibration
rather than statistical noise.  One major source of systematic error in
determining SNe Ia color evolution (needed for distance estimation) is
uncertainty in telescope transmission, both within and between surveys.  We
introduce here the Collimated Beam Projector (CBP), which is meant to measure a
telescope transmission with collimated light. The collimated beam more closely
mimics a stellar wavefront as compared to flat-field based instruments, allowing
for more precise handling of systematic errors such as those from ghosting and
filter angle-of-incidence dependence.  As a proof of concept, we present CBP
measurements of the \sd prototype telescope, achieving a standard ($1\sigma$)
uncertainty of $3$~\% on average over the full wavelength range measured with a
single beam illumination. 
\end{abstract}


{\noindent \footnotesize\textbf{*}Nicholas Mondrik,  \linkable{nmondrik@gmail.com} }

\begin{spacing}{1}

\section{Introduction}
\label{sec:intro}
Multi-band photometry permits measurements of much fainter sources than
spectroscopy while still preserving low spectral resolution components of the
observed SED (spectral energy distribution).  In general, there is a large number
of desirable photometric measurements that are useful to astronomers, including,
but not limited to: top-of-atmosphere (TOA) flux, flux corrected for Galactic
extinction, and flux corrected for both Galactic and host-galaxy extinction, for
supernova (SNe) cosmology in particular\cite{Bessell2005, Scolnic2014}.
Only instrumental magnitudes are directly measured, and include
contributions from telescope transmission idiosyncrasies and atmospheric
transmission variations that serve to obscure astrophysically interesting
features\cite{Stubbs2006}.

Knowledge of instrumental passbands is particularly useful when attempting to
determine magnitudes of sources having SEDs that are dissimilar to standard flux
calibration stars, as in the case of SNe Ia cosmology.  With the advent of
large-scale transient surveys such as Rubin Legacy Survey of Space and
Time\cite{lsstSRD,lsstdataprod} (LSST) undertaken by Rubin Observatory,
the number of Type Ia supernovae will be large enough that systematic
calibration uncertainty will become the limiting factor in the determination of
cosmological parameters\cite{Stubbs2015}.  Additionally, state of the art survey
calibration schemes, such as the Forward Global Calibration
Method\cite{Burke2017}, can take passband measurements as inputs, thereby
increasing the accuracy and precision of survey measurements by accounting for
the variations in the spectra of field stars.

The goal of photometric calibration is to arrive at a measure of brightness that
is an accurate measurement of the source SED integrated over a given
bandpass and that accounts for temporal variations in the atmospheric and
optical transmission functions.  The material presented here is focused on the
problem of determining the optical transmission function of the telescope, but
an overview of factors impacting photometric calibration of surveys can be found
in Ref.~\citenum{Li2016}.  By optical transmission function, we mean the
fraction of photons that enter the telescope, are converted to photo-electrons,
and subsequently read out of the CCD (charge-coupled device).  The simplest
method of measuring this quantity would then be to send a known number of
photons down the telescope's optics, and compare the number measured on the CCD
to the number originally emitted.

The usual way to accomplish this mapping in astronomy is by tying back to almost
pure hydrogen white dwarfs stars, for which we believe we are able to estimate
the flux based on spectral line measurements and radiative transfer calculations
\cite{Bohlin2014}. Tying to standards that can actually be built on Earth has
also been tried, either by creating a proxy of a black body that can be directly
observed by the telescope \cite{Hayes1975}, or by creating a calibrated stable
light source using a calibrated detector, usually a photodiode provided by the
National Institute of Standards and Technology (NIST)\cite{Stubbs2006}. While
this latter approach relies on a metrology chain with more intermediate steps,
it capitalizes on the fact that NIST photodiode's quantum efficiency (QE) can be
calibrated with an uncertainty on the order of $0.1 \%$ over the full optical
range (400nm-1000nm).

This is the approach we choose for the Collimated Beam Projector: By using a
NIST-calibrated photodiode to normalize away the variations in the calibration
light source, we are seeking to transfer a known flux scale, such as the one
defined by the Primary Optical Watt Radiometer\cite{Houston2006} (POWR) at NIST,
onto a telescope CCD which can then transfer that calibration to an
astrophysical source.  Technically speaking, POWR provides an optical-watt
\textit{power} scale, which is distinct from a \textit{flux} scale by a factor
of area (Flux $\equiv$ Power/Area).  Because we are interested here only in
chromatic variations, and not absolute calibration, we may treat the two as
equivalent.  Although an \textit{absolute} flux scale is desirable (transmission
of the system is known exactly at all wavelengths), in many practical cases a
\textit{relative} flux scale is sufficient.  By relative flux scale, we mean
that the transmission ratio $T(\lambda)/T(\lambda')$ is known between any two
generic wavelengths, but the overall grayscale normalization is unknown.  Said
more simply, we need to ensure that there are no temporal or spatial (over the
detector) variations in observed ratios of fluxes (colors), and that these
ratios are consistent with the true flux ratios.

\subsection{Extant transmission measurement devices}
Astronomers have wrangled with the challenges of calibrating CCD-based
observations since the advent of the devices in the late 20$^{\mathrm{th}}$
century.  Bias frames, flat field exposures, star flats, and many other types of
data are frequently used by astronomers to contend with the temporal and spatial
non-uniformity of telescope optical systems.  The flat field is of particular
interest to the challenge of determining instrumental passbands because it is an
attempt to standardize the response of each pixel in a telescope system.  A
typical flat field system uses a white-light source to illuminate a lambertian
reflecting screen, usually far out-of-focus, which is observed by the telescope,
with the end result being a ``flat'' (constant surface brightness) image on the
detector.  The flat field obtained, which has different scattered and stray
light behavior than a start field science beam, generally tends to homogenize
variability arising from the screen construction, and is used to normalize away
pixel-to-pixel variations in images of astrophysical sources.

However, as has been pointed out (e.g., Refs.~\citenum{Stubbs2014, Baumer2017}),
naive application of flat fields may introduce systematic errors that limit the
ultimate uncertainty of the measurement. A flat field may not perfectly mimic
the response to astronomical sources due to several factors such as delineating
the difference between variations in pixel size, caused by departure from a
rectilinear grid (due, for example, to lateral electric fields within the sensor
that distort pixel gridlines) from true variations in pixel QE.  One of the
fundamental issues with conflating these two processes is that variations in
pixel size are flux-conserving (although a given photo-electron may end up in a
neighboring pixel, it is still present and can be counted), while variations in
QE are not (the photo-electron is never generated in the first place).

The sheer ubiquity of flat field screens at observatories does however, make
them a tantalizing component to leverage in the challenge of measuring telescope
throughputs.  It remains then, to concoct a scheme by which variations in pixel
size can be decoupled from variations in QE.  One method is to use tunable,
narrowband light to illuminate a flat field screen, generating a data cube of
flats at each wavelength over the spectral range of interest.  By looking at the
signal variation of each pixel with wavelength, one can remain agnostic to the
\textit{average} size of the pixel.  We stress \emph{average}, because the
method is still sensitive to the wavelength-dependent component of pixel size
variations.  As an example, in the presence of a static lateral electric field
(for example, from impurity gradients), a photo-electron from a blue photon,
which converts near the back surface of the CCD, will experience greater
deflection than a red photon, which converts deeper in the device.  This would
make the pixel in question appear less responsive in the blue than the red, and
again the flux-conserving pixel size variations can be mistaken for QE
variation.  Overall, this type of method reduces confusion between pixel size
and QE, becoming instead limited by the wavelength-derivative of pixel size
variations.  More appropriately, they might be said to be limited by the
\textit{differential} size-derivative of pixels, since leakage of
photo-electrons out of a pixel can be compensated for by leakage into the pixel
from its neighbors.

Several experiments and devices have been constructed along these lines, for
example, the method proposed in Ref.~\citenum{Stubbs2006} and implemented in
Ref.~\citenum{Stubbs2010} used a photodiode-monitored, monochromatic, tunable
laser source reflected off of a flat field screen to measure the transmission
function of the PanSTARRS telescope.  The DECal system\cite{Rheault2012,
  Marshall2016} on the Cerro Tololo Inter-American Observatory's Blanco
telescope, similarly illuminates a flat field screen with a combination of LEDs
and a white-light powered monochromator, which is monitored by a photodiode.

Challenges remain for flat-field based systems, however. In particular, there
are systematic differences in scattered light paths between flat field and
stellar illumination patterns. These differences would result in systematic
errors when deriving transmission measurements for point sources.  It would be
ideal, then, to illuminate the telescope with a wavefront similar to that of a
star: i.e., full-pupil and planar.  This would effectively side-step the
challenging task of deriving corrections for point-source images from
surface-brightness based flats.

The SCALA system\cite{Lombardo2017} on the University of Hawaii 88-inch
Telescope uses a white-light powered monochromator as its illumination source,
which is fed into a series of integrating spheres attached to a frame mounted on
the interior of the telescope dome.
Apertures on the outputs of the integrating spheres are collimated by mirrors and
re-imaged onto the focal plane of the SuperNova Integral Field Spectrograph
(SNIFS). Two of the collimated beams is monitored by a Cooled Large Area
Photodiode\cite{Regnault2015}, which provides normalization.  In this case, full
pupil illumination is traded off against using a collimated beam re-imaging
system to project large (1$^\circ$) spots onto the focal plane. Tying the
calibration of SNIFS to the NIST definition of the optical Watt in turn permits
the SNfactory to provide, by repeated observations of standard stars over more
than a decade, a well calibrated star network covering a large fraction of the
sky observable from Hawaii\cite{Rubin2xx}.

The NIST Stars project uses a spectrograph to observe a NIST-calibrated light 
source and standard stars alternately, thereby transferring the light source
calibration to the standard
stars\footnote{https://www.nist.gov/programs-projects/nist-stars}.  When the
light source is placed sufficiently far away, the wavefront is effectively
parallel when entering the telescope, thus generating a full-pupil collimated
beam.

The StarDICE project is of the same flavor, but uses a stable calibrated
poly-chromatic source made of LEDs placed at $\sim 200$~m from a $\sim 1$~m
focal length telescope as an artificial star. It aims at providing a network of
stars of magnitude between 10 and 13 calibrated with traceability to the SI
(Système International d'Unités) through the NIST POWR scale by observing in
turn the artificial star and the CALSPEC stars. This procedure makes the
measurement of the transmission of the StarDICE telescope mandatory.

In this paper, we present the Collimated Beam Projector (CBP) instrument, a
telescope throughput measurement system\cite{Coughlin2016, Coughlin2018,
  Ingraham2016}.  In Sec.~\ref{sec:DesignConsiderations}, we explore design
considerations in general, and in Sec.~\ref{sec:CBPDesign} we describe the
design and components of our CBP system in particular.
Section~\ref{sec:CBPCalib} outlines the method for tying CBP measurements back
to the flux scale established at NIST, Sec.~\ref{sec:CBPExperimentSetup}
describes the \sd experiment and CBP setup, and Sec.~\ref{sec:CBPReduction}
describes the data reduction procedure.  Section~\ref{sec:CBPResults} presents
transmission curves for the \sd prototype telescope, taken at the Laboratoire de
Physique Nucl\'{e}aire et des Hautes \'{E}nergies (LPNHE), and presents lessons
learned during this phase.  Finally Sec.~\ref{sec:Conclusion} reviews planned
upgrades and revisions to the system.


\section{Design Considerations for Optical Transmission Measurement}
\label{sec:DesignConsiderations}
To design an instrument to measure the optical throughput of an imaging system,
we must first understand both the optical properties and measurement goals of
the system. Using $t$ for time, $\lambda$ for wavelength, $\mathbf{x}$ for the
vector position, $\boldsymbol{\gamma}$ for a given altitude and azimuth, and
$\mathbf{r}$ the location on the primary mirror, the flux arriving from a given
source on the focal plane, $\phi(t,\mathbf{x},\boldsymbol{\gamma})$, can be
expressed as

\begin{equation}
\label{eqn:CCDsignal}
    \phi(t,\mathbf{x}, \gamma) = C(t)\int A(t,\lambda,\boldsymbol{\gamma}) \, T(t,\lambda,\mathbf{x},\mathbf{r},\boldsymbol{\alpha}) \,
    S(t,\lambda, \boldsymbol{\alpha}) \,
    d\lambda \, d^2\boldsymbol{\alpha} \, d^2\mathbf{r},
\end{equation}

where $C(t)$ is a constant with respect to focal plane position, telescope
pointing, and wavelength (e.g., collecting area, electronic gain), $A(t,
\lambda, \boldsymbol{\gamma})$ is the instantaneous atmospheric transmission
function, $T(t, \lambda, \textbf{x}, \textbf{r}, \boldsymbol{\alpha})$ is the
instrumental transmission function, and $S(t, \lambda, \boldsymbol{\alpha})$ is
the total TOA photon flux incident on the primary from sources at angle
$\boldsymbol{\alpha}$ relative to the pointing of the telescope.  The integrals
are over all wavelengths, relative angles, and the entirety of the primary,
respectively.

To measure the transmission of a telescope, we must contend with all of the
terms above, with a few modifications. Relative to standard astronomical
measurements, $A$ is much suppressed due to shorter path-lengths and $S$ is
instead the spectrum of the calibration light.  Our ultimate goal is to estimate
$T$, the transmission function of the telescope imposed on astrophysical
point-source wavefronts.  Armed with this expression, we can begin to explore
the design requirements for our transmission measurement systems.

\subsection{Motivation for using collimated beams}

A planar wavefront incident on the primary at a defined angle
$\boldsymbol{\alpha}$ has a fixed ghosting pattern as a function of wavelength
set by the geometry of a given optical system.  When a telescope is illuminated
with a non-planar wavefront, ghosts and other scattered light are superimposed
on the target region, resulting in a subtly different transmission function than
the one experienced by an astrophysical source.  For this reason, flat fields,
which illuminate the telescope at all angles, are not ideal calibrators for
point sources.

Beyond the angle of incidence, the location of a photon's impact on the
telescope primary, $\mathbf{r}$, can play a major role in transmission
measurements, particularly for interference filter-based systems that are not
designed with variable multi-layer coating thicknesses to account for
geometrically-related shifts in bandpasses.  For these systems, the incidence
angle $\theta$ at which the light passes through the filter changes the
effective transmission of the filter.  An approximation of the shift in
transmission for a filter for a given incidence angle $\theta$ and index of
refraction $n$ is\cite{Regnault2009}

\begin{equation}
    T(\lambda, \, \theta) = T\Big(\lambda \, ( 1 - \frac{\sin ^2{\theta}}{n^2}) ^{-\frac{1}{2}}, \, \theta = 0 \Big).
\end{equation}
where we see that the effect is essentially a blueshift of the filter
transmission at $\theta=0$.

If the wavefront injected into the telescope is non-planar, there will be a
difference in the measured filter transmission and transmission function
experienced by a star, the scope of which is dependent on telescope geometry (in
particular, the measured filter passband will be broadened according to the
above equation, weighted by the relative photon flux density at each angle
$\theta$).

The location of filter edges is of particular concern, so it is worth
understanding their impact on transmission measurement devices.  Photons with a
wavelength located within an edge are highly likely to experience internal
reflection within the filter because the transmission significantly deviates
from unity and from zero, by definition.  These internal reflections can escape
the filter and land on the detector, contributing significantly to scattered
light.  For a collimated system, there is some hope that these scattering
patterns approximate (to a degree) that of a stellar wavefront.  For a flat
field system, whose photon phase-space distribution is different than that of
a plane wave illumination (with a single phase and direction), there is additional
systematic uncertainty.

Together, these two concerns motivate the use of collimated light, rather than
flat fields.  It is important to note that here we have assumed that the goal of
the survey is to measure point sources -- if measurements of surface brightness
are required, then projecting a planar wavefront onto the telescope is not a
requirement anymore, and flat-field based methods are more appropriate.

\subsection{Light source requirements}
A first requirement is based on the tautology that the light source used to
measure the transmission function must be capable of emitting light over the
entire wavelength range of interest (roughly 300 nm - 1100 nm for optical
systems).  Assuming negligible attenuation from atmospheric effects ($A$ is
small in setups where the light source is close to the telescope), the relevant
wavelength-dependent transmission variations to be measured are those in $T$.
This motivates an additional property of the light source: that the optical
bandwidth should be small compared to variations expected in $T$.  For
interference filter based systems, this scale is set naturally by the filter
edges, which transition from opaque to transparent over roughly a 20 nm span.
This results in a filter transmission change ratio of around 1\% per $0.2$ nm.
Properly sampling smoothly if rapidly varying filter edges at the percent level
therefore suggests optical bandwidths of order 1 nm, which is readily achievable
by devices such as monochromators and tunable lasers.  In addition to the
wavelength requirement, the output flux of the light source, $S$, must be known
at the sub-percent level.

\section{Design of the Collimated Beam Projector (CBP)}
\label{sec:CBPDesign}
We present here an overview of the CBP system; see
Refs.~\citenum{Coughlin2016,Coughlin2018} for additional details on CBP design.

The CBP consists of two components: an imaging system and a
light source.  A tunable laser (with associated coupling optics) provides
high-power, narrowband light for the CBP imaging system.  The imaging system is
composed of a collimating optic, a focal plane mask, and an integrating sphere,
with a fiber optic cable connecting the output of the laser to the integrating
sphere.  This system is then attached to an alt-az mount that allows the CBP to
be pointed remotely.  By coordinating the alt-az pointing of the CBP and the
telescope, it is possible to point the CBP beam at a different section of the
primary while leaving the position on the detector fixed.  This process, which
we dub ``pupil stitching'', allows the CBP to scan the full primary mirror (and
by extension, the different paths taken by photons en route to the same detector
pixels).  This in theory enables a synthetic full-pupil measurement with a
collimated beam (essentially emulating a stellar wavefront). Given hard time
constraints in particular due to the limited availability of the laser at LPNHE,
we defer the full pupil measurement to a forthcoming paper (Souverin et 
al. in prep.).

Taking the constraints imposed above and re-writing Eqn.~\ref{eqn:CCDsignal} for
an input photon distribution provided a single CBP pointing, we obtain

\begin{equation}
\label{eqn:CBPsignal}
    \phi_\mathrm{CBP}(t,\textbf{x},\lambda) = C(t) \int_\mathrm{CBP} T(t,\lambda,\textbf{x},\textbf{r},\boldsymbol{\alpha}) \, 
    S(t,\lambda, \boldsymbol{\alpha}) \, d^2\textbf{r},
\end{equation}

where the integral is over the CBP's footprint on the primary, and we have
assumed the CBP's output flux $S(t,\lambda,\boldsymbol{\alpha})$ can be
described as $S(t, \lambda', \boldsymbol{\alpha'}) \, \delta(\lambda'-\lambda)
\, \delta(\boldsymbol{\alpha}' - \boldsymbol{\alpha})$, i.e. the CBP output is
monochromatic with a well-defined angle relative to the telescope (indicated by
the Dirac delta functions $\delta (\dots)$), and that atmospheric attenuation is
negligible.  In the case where the CBP's footprint is small relative to the
primary, one can additionally multiply by another factor of
$\delta(\textbf{r}'-\textbf{r})$, and assume that the measurement is a
point-like sampling of the primary.  Alternatively, one might make the assertion
that the system's optical transmission is not a strong function of the input
beam's location on the primary, and thereby absorb the (now constant) integral
into $C(t)$.  In all generality, this latter assertion is untrue since different
input beam locations on the primary result usually in different angles of the
converging beam passing through an interference filter before hitting the
detector, though the level of deviation is dependent on telescope geometry and
is generally smaller for slower optical systems, which operate at smaller
angles.

This assumption leaves for a later work (Souverin et al. in prep) the discussion
about the fact that different areas on a primary mirror can have very different
reflectivity properties. The actual measurement of this variability will be part
of the strategy developped in forthcomming papers (StarDICE collaboration
2023-2024) in order to reconstruct the full pupil transmission from CBP patch
measurements.

\subsection{Tunable Laser}
The light source for the CBP is an EKSPLA NT-242 tunable
laser\footnote{Identification of commercial equipment to specify adequately an
  experimental problem does not imply recommendation or endorsement by the NIST,
  nor does it imply that the equipment identified is necessarily the best
  available for the purpose.  This applies for all other commercial products
  named in this publication.}, which outputs 3 ns to 6 ns laser pulses at a 1
kHz repetition rate, with a total output power of 0.5 W at roughly 450 nm.  The
tunable laser uses a non-linear optical process (spontaneous parametric
down-conversion, SPDC) inside of a crystal (called an optical parametric
oscillator, or OPO) to convert an incident photon of frequency $\nu_1$ (the OPO
pump beam) into two photons such that $\nu_1 = \nu_2 + \nu_3$, where energy and
momentum remain conserved.  In the case of the NT-242, these two photons are
cross-polarized (formally, this means the process is type-II SPDC), which allows
for separation of the ``signal'' and ``idler'' beams (the high- and
low-frequency output beams, respectively) via a Rochon prism within the laser.
In the case of the NT-242 laser used here, the pump beam is provided by a Nd:YAG
laser at 1064 nm.  This pump beam is tripled in frequency to 355 nm before
reaching the OPO, meaning the OPO itself is pumped by a 355 nm beam, which
allows access to wavelengths above 355 nm.  In order to reach wavelengths below
the OPO pump wavelength, the beam is sent through a second harmonic generator
(SHG), which doubles the frequency of the incoming light.  The SHG allows access
to wavelengths below 355 nm, at the cost of much-reduced efficiency.  In the
end, the NT-242 laser is tunable from below 300 nm to over 2 $\mu$m, which is
well matched to the sensitivity range of CCDs.  The tunable laser provides high
flux in a narrow bandpass, $\lesssim 5$ cm$^{-1}$, which corresponds to
approximately $\delta\lambda=0.13$ nm at 500 nm, and $\delta\lambda=0.5$ nm at 1
$\mu$m.  These are upper limits quoted by the manufacturer, and measurements
taken using these systems show bandwidths of 0.08 nm to 0.48 nm between 350 nm
and 1100 nm\cite{Woodward2018}.  As these bandpasses are small compared to the
accuracy achieved at this stage, we treat the output of the laser as
monochromatic.

There are additional concerns that must be addressed when using tunable lasers,
and we present here the major challenges; for an overview of tunable lasers in
optical calibration applications, see Ref.~\citenum{Woodward2018}.  There are
three primary obstacles to overcome: excessive brightness in some regions, low
efficiency near the degeneracy point of the system, and incomplete separation of
the signal and idler\footnote{We remind here the \emph{idler beam} refers to the
  second beam obtained after wavelength splitting by the OPO, the \emph{signal
    beam} denoting the beam at the desired wavelength.} beams in the degeneracy
region (around 710 nm for the EKSPLA NT-242).  To address brightness concerns,
we added a reflective neutral density (ND) filter to the fiber coupling system so
that the beam can be attenuated.  The degeneracy region occurs when
$\nu_\mathrm{signal} \simeq \nu_\mathrm{idler} \simeq \nu_\mathrm{pump}/2$, and
is characterized by low output power and poor separation of the signal and idler
beams.  Low efficiency in the degnereracy region can be overcome either by
increasing integration times, or by using a different light source.  For
example, an OPO pumped at 532 nm would have its degeneracy point at 1064 nm, in
a part of the spectrum farther from filter edges.  Enhanced separation of the
signal and idler beams can be achieved in several ways, for example, with the
addition of rotating polarizers or short-/long-pass filters with cut-off/-on
wavelengths equal to the OPO pump beam wavelength.  Such optical elements are
not present in our current fiber coupling scheme, but will be installed for our
next iteration.

\begin{figure}[H]
\begin{center}
\begin{tabular}{c}
\includegraphics[width=\textwidth]{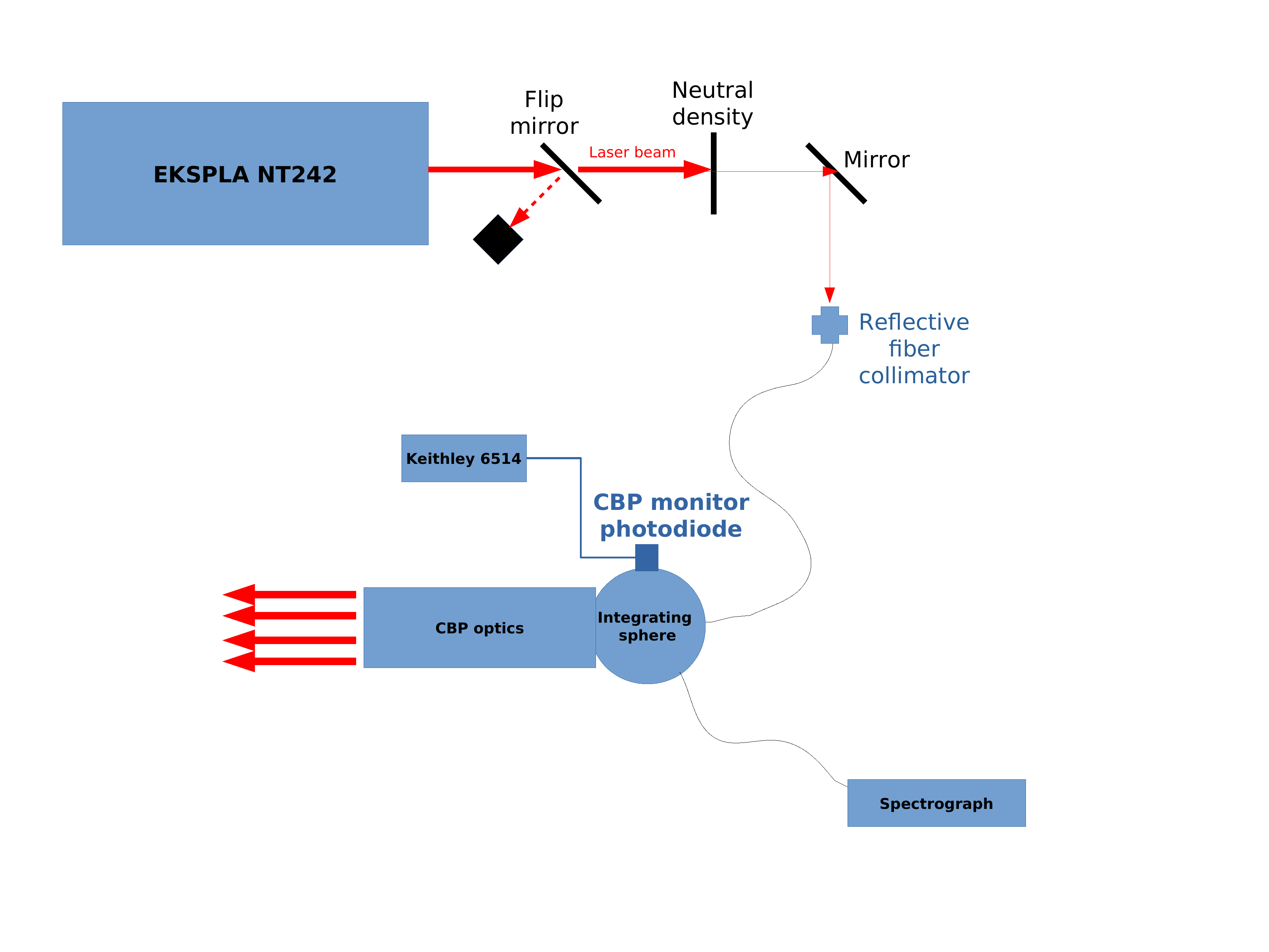}
\end{tabular}
\end{center}
\caption 
{  \label{fig:CBPsketch}
Schematic of the CBP light injection. The telescope measured is placed in front
of the CBP optics. 
}\end{figure}

The light first passes through a flip mirror, which optionally steers the beam
into a beam dump, shuttering the laser.  Afterwards, the beam passes through a
variable ND filter mounted on a rotation stage.  This filter allows for
attenuation of the beam in regions where the laser is so bright as to require
sub-second exposure times.  The beam then is coupled into a reflective fiber
collimator, which is connected to the integrating sphere via an optical
fiber. Figure \ref{fig:CBPsketch} summarizes this setup.

\subsection{CBP imaging system}
\label{subsec:CBPImaging}
The CBP imaging system is comprised of a
Sonnar\textsuperscript{\tiny\textregistered} CFE Superachromat 5.6/250 mm lens,
which images a mask held by a Finger Lakes Instrumentation (FLI) CL1-10 filter
wheel backlit by a Labsphere integrating sphere.  The lens' focus is not
strongly chromatic, which should allow us to forgo re-focusing the CBP optics at
each wavelength.  The FLI filter wheel contains two internal wheels; one holds
the masks to be re-imaged, and another holds an f-stop aperture, which prevents
injection of light at angles too extreme to be focused by the lens.  The CBP
currently holds 3 imaging masks: a 20 $\mu$m pinhole, a 5x5 grid of 20 $\mu$m
pinholes at 200 $\mu$m spacing, and a large 500 $\mu$m pinhole.  The single 20
$\mu$m allows for precise determination of ghost locations without confusion
from nearby pinholes, while the pinhole grid allows for multiplexing of
transmission measurements as a function of sensor location.  Using the CBP with
no pinhole allows for coarse alignment of the CBP and telescope, while the 500
$\mu$m pinhole is useful in achieving fine alignment and for providing a
resolved, locally flat region on the detector.  The 500 $\mu$m pinhole is also
required for calibrating the CBP, as the smaller pinholes do not provide
sufficient brightness.  It should be noted that such small pinholes tend to have
large variations in their diameters, and the manufacturer of the masks used
here, Lenox Laser, quotes $\pm 10$ \% tolerance on the diameter, which implies
variations in pinhole area of up to 50 \% in the worst-case scenario.

In addition to the output port and light injection port, there are two devices
attached to the integrating sphere.  The first is the monitor photodiode, a
Thorlabs SM1PD2A, which normalizes away temporal variations in laser power.  The
photodiode is connected to a Keithley 6514 charge-integrating electrometer,
which measures the amount of charge collected by the photodiode during an
exposure.  Because there is a saturation limit to the collection capacity of the
electrometer's measurement capacitor, an aperture of 1 mm is placed in front of
the photodiode, allowing for use of integration times commensurate with those
needed for telescope measurements.  The second device connected to the
integrating sphere is an Ocean Optics QE65000 fiber-fed universal serial bus
spectrograph, which monitors the wavelength emitted by the laser.  The
integrating sphere itself is necessary to ensure that all pinholes are
illuminated homogenously and achromatically with respect to the spectrograph and
the monitoring photodiode, since we cannot monitor the flux
emitted by each individual pinhole.  If the surface brightness seen by the
pinhole grid varied with wavelength across the grid, it would imprint itself as
a focal plane transmission gradient on the telescope.  The integrating sphere
also ensures that the light seen by the monitoring equipment (photodiode,
spectrograph) has the same surface brightness as the light illuminating the
pinhole grid. It has to be noted that these desirable features come at the price
of decreasing the surface brightness of the pinholes in direct proportion of the
size of the integrating sphere used. In the current design, the integrating sphere
diameter (6 inches) has been fixed by the desire to multiplex the transmission measurement
using a grid of multiple pinholes. This in turn resulted in a large flux
dillution that required such a powerful light source as the laser. 

\begin{figure}[H]
\begin{center}
\begin{tabular}{c}
\includegraphics[width=\textwidth]{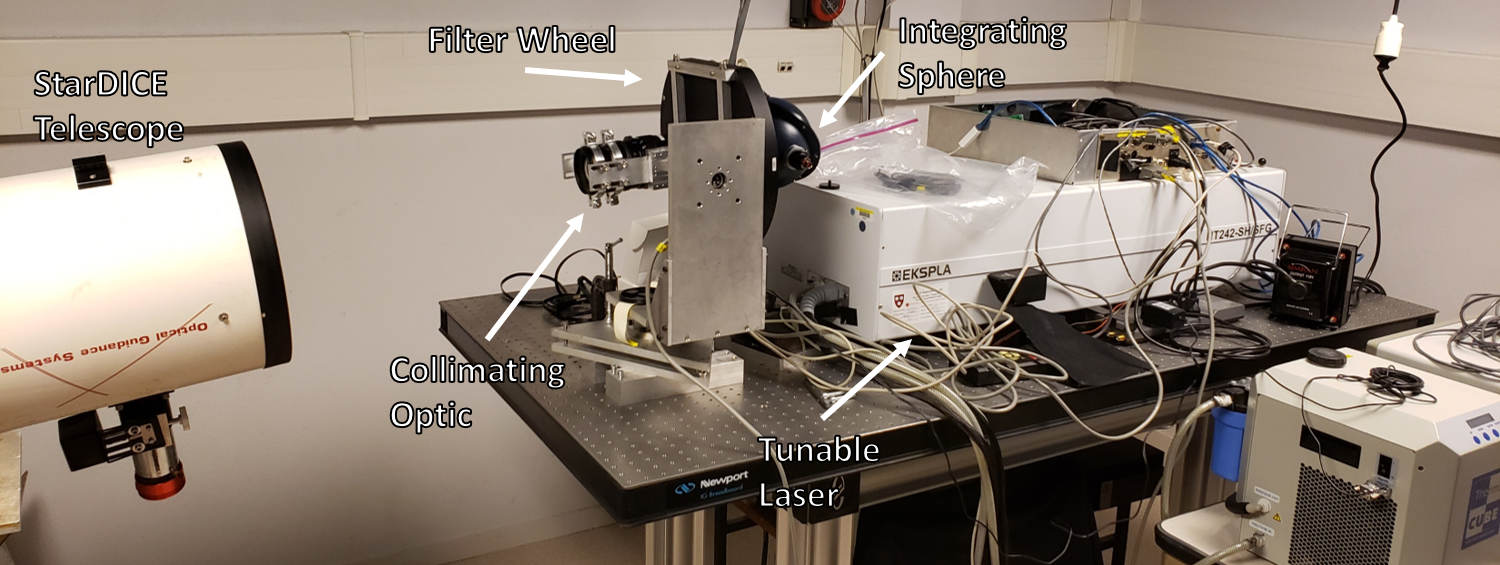}
\end{tabular}
\end{center}
\caption 
{  \label{fig:CBPPic}
An image of the CBP installed in the StarDICE lab at LPNHE.
}\end{figure} 

\section{Establishing the CBP flux scale}
\label{sec:CBPCalib}
In order to transfer the detector-based flux scale established at NIST to the
CBP, the CBP optics were sent to NIST to be calibrated against one of the
trap detectors used to hold the POWR optical watt scale.  This step is necessary
because the CBP input flux is monitored inside the integrating sphere, but the
light must pass through an additional strongly chromatic element (the
collimating lens) prior to entering the telescope.  It is therefore necessary to
have an additional calibration at each wavelength of interest between the
integrating sphere flux measured by the CBP monitoring photodiode and the actual
light emitted by the system.

\begin{figure}
\begin{center}
\begin{tabular}{c}
\includegraphics[width=\textwidth]{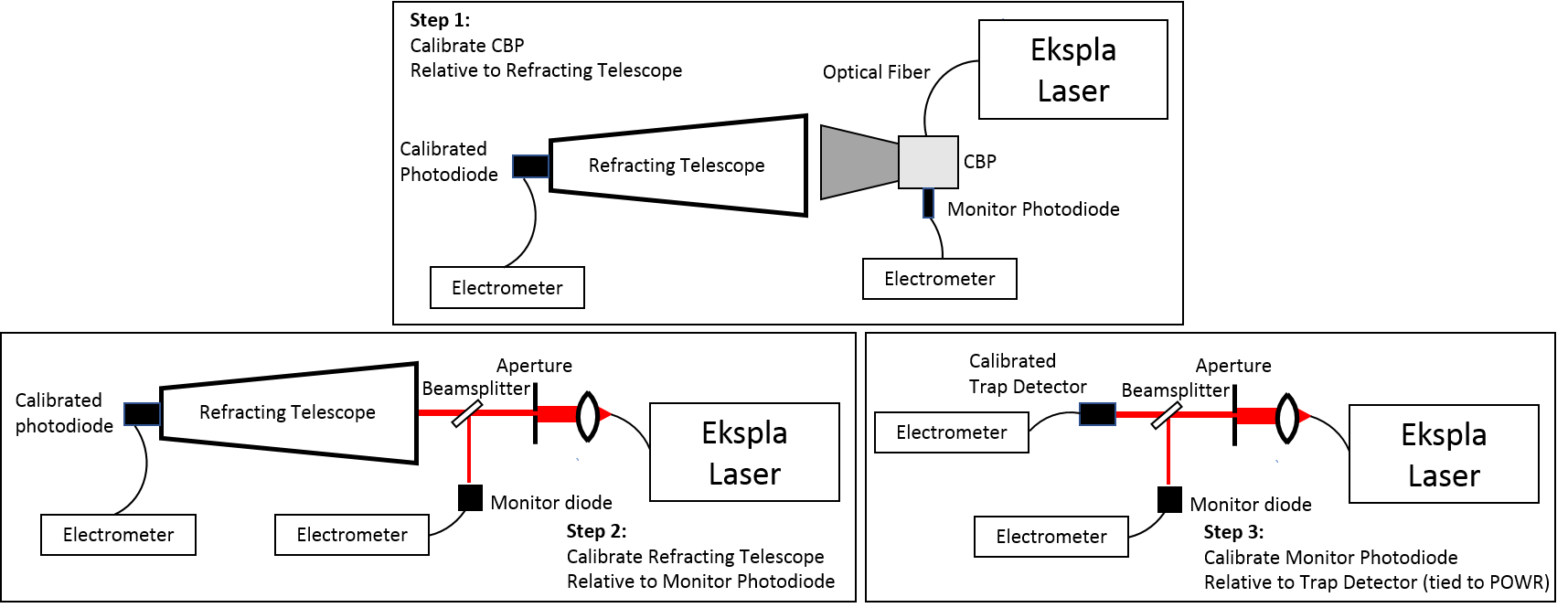}
\end{tabular}
\end{center}
\caption 
{ \label{fig:CBPCalibScheme} A schematic illustrating the three-step process
  used to transfer the POWR-traceable calibration of a NIST trap detector to the
  CBP.  Step 1 measures the ratio $Q_\mathrm{CBP}/Q_\mathrm{Telescope}$ measured
  respectively by the CBP monitor photodiode and the NIST calibrated
  photodiode. Likewise the ratios $Q_\mathrm{Telescope}/Q_\mathrm{Monitor}$ and
  $Q_\mathrm{Monitor}/Q_\mathrm{Trap}$ for steps 2 and 3 are obtained by
  comparing the measurements of the Monitor diode with the NIST calibrated
  photodiode and the Calibrated Trap detector respectively.  The product of the
  ratios, multiplied by the responsivity of the trap detector, gives the
  responsivity of the CBP in units of [A W$^{-1}$].  The inverse of the CBP
  responsivity is defined to be the CBP calibration factor. The calibrated
  photodiode refered to in step 1 and 2 is a NIST-calibrated photodiode with
  respect to POWR.  }\end{figure}

This calibration took place in three steps. An overview of the calibration
process is shown in Fig.~\ref{fig:CBPCalibScheme}. The reason why the CBP was
not directly calibrated using the calibrated photodiode is because the
calibrated photodiode alone is too small to sample the entire CBP output beam,
and had poor signal-to-noise when only subsampling the beam. A different
calibration scheme, using larger solar cells tied to the NIST optical watt
definition\cite{Brownsberger2021} is currently being implemented and will be
presented in a forthcoming publication (Souverin et al. in prep.).

During all steps, all optics and photodiodes were in a light-tight box, and the
laser was coupled into the box through a fiber-optic cable.  During each step
of measurement, the laser was tuned automatically across the spectral range for
calibration.  Because the laser delivers light pulses of $\sim 10~ns$ at a
frequency of 1~kHz, we choose to operate the Keithley 6514 in charge accumulation
mode.  For this mode of operation, a computer-controlled
shutter at the laser output limits the exposure time (typically 1 sec.) of the
laser to the fiber.  The two photodiodes (the CBP monitor photodiode and the
calibrated photodiode, see below) were connected to a separate
charge accumulation electrometer, and both electrometers started charge
accumulation just before the opening of the laser shutter and stopped
accumulation just after closing of the laser shutter.  Subsequently, the total
amount of accumulated charge, in Coulombs, was read out from each electrometer.

\begin{figure}
\begin{center}
\begin{tabular}{c}
\includegraphics[width=\textwidth]{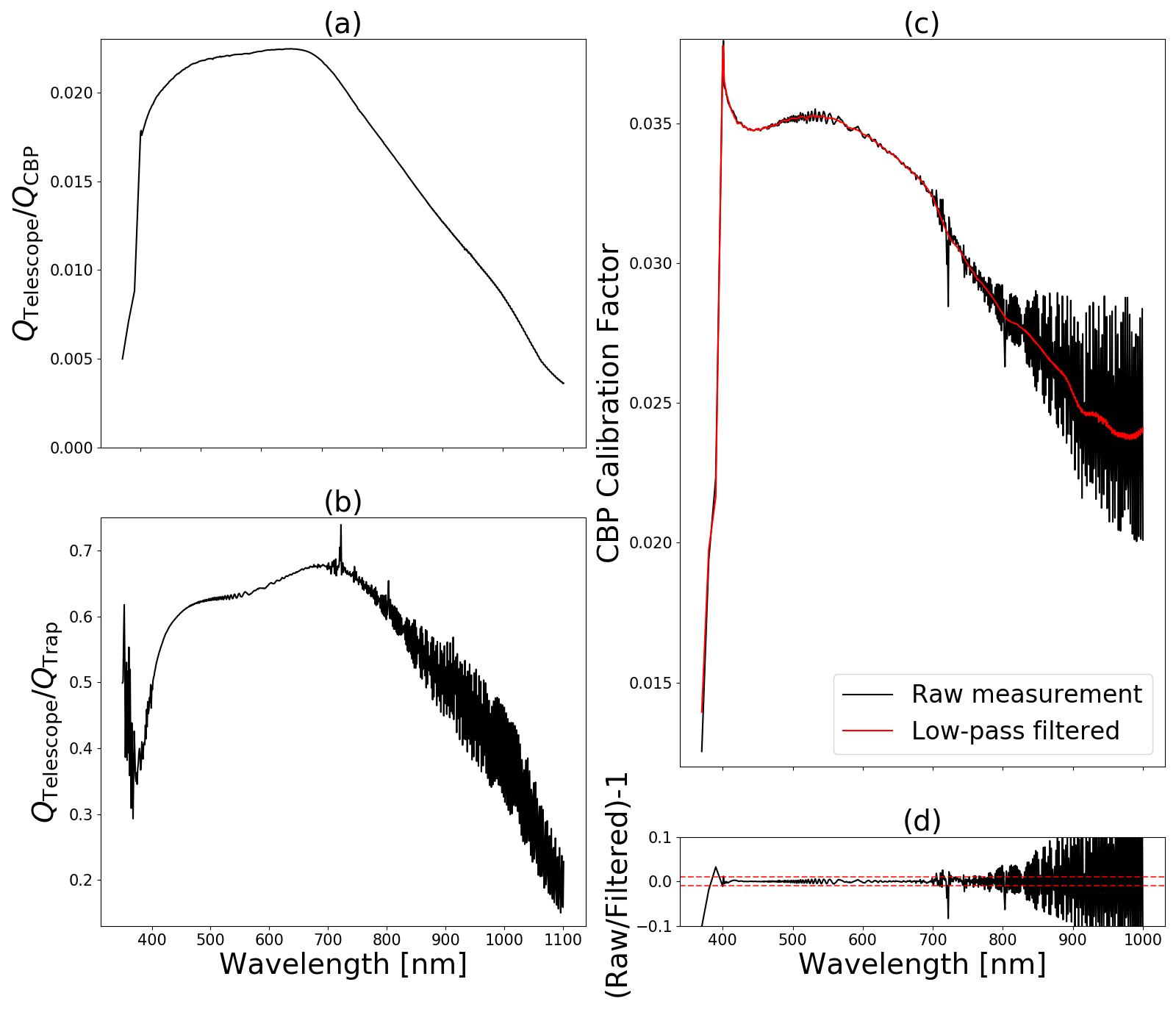}
\end{tabular}
\end{center}
\caption 
{ \label{fig:CBPCalib} Results from the calibration chain described in
  Sec.~\ref{sec:CBPCalib}.  (a): Ratio between calibrated photodiode response and
  CBP monitor photodiode response.  (b): Ratio between calibrated photodiode
  response and trap detector response.  (c): The calibration factor of the CBP.
  The red curve has been low-pass filtered to suppress the interference fringe
  signal from the calibration telescope-trap measurement.  (d): The relative
  difference between the raw and low-pass filtered CBP system throughputs, with
  dashed red lines drawn at $+/- 1$ \%.  The sharp cutoff at approximately 400
  nm seen in (a) and (c) is attributed to the AR coating on the CBP lens.  The
  residuals in (d) are used to estimate the systematic CBP calibration
  uncertainty due to fringing.}
\end{figure}

In step 1, the CBP illuminated the entrance pupil of a 100 mm diameter
refracting telescope with a NIST-calibrated photodiode (referred to as the
``calibrated photodiode'' in the following and in step 1 and 2 of
Fig.~\ref{fig:CBPCalibScheme}) at its focus.  The focused beam underfilled the
calibrated photodiode, ensuring the full collection of the incoming beam.  The
charge $Q_\mathrm{CBP}$ on the CBP monitor photodiode, and the charge
$Q_\mathrm{Telescope}$ on the calibrated photodiode, were read out after one
laser exposure cycle at each wavelength.  The ratio
$Q_\mathrm{CBP}/Q_\mathrm{Telescope}$ then references the CBP photodiode to the
calibrated photodiode.  Data for this step, plotted as
$Q_\mathrm{Telescope}/Q_\mathrm{CBP}$, is shown in Fig.~\ref{fig:CBPCalib}a.

The next two steps reference the calibrated photodiode to the POWR-calibrated
trap detector through an intermediate monitor photodiode.  The trap detector
used here is a 3-element arrangement of photodiodes such that a plane-parallel
beam of incoming light must make 5 reflections before escaping, boosting the
effective quantum efficiency of the trap relative to a single photodiode.  These
two steps use a different fiber optic cable than the first step, and the fiber
optic transmitting the laser light was passed through a speckle reducer that
modulated a section of the fiber at high frequency to mix the modes.  Within the
light-tight box, the fiber output port was placed at the focus of a 100 mm,
f/2.8 refractive collimator.  The resulting collimated beam was passed first
through an aperture to reduce its diameter to 5 mm, then through a beamsplitter
with an intermediate monitor photodiode (10 mm x 10 mm called monitor diode in
Fig.~\ref{fig:CBPCalib}a-c) set to capture the reflected beam.

For step 2, the beam transmitted through the
beamsplitter illuminated the refracting telescope with the calibrated photodiode
at its focus.  The charge $Q_\mathrm{Telescope}$ on the calibrated photodiode and
the charge $Q_\mathrm{Monitor}$ on the intermediate monitor photodiode were read
out at each wavelength, so the ratio $Q_\mathrm{Telescope}/Q_\mathrm{Monitor}$
references the calibrated photodiode to the intermediate monitor.

For the step 3, the beam transmitted through the beamsplitter illuminated and
completely underfilled the trap detector.  The charge $Q_\mathrm{Monitor}$ on
the intermediate monitor photodiode and the charge $Q_\mathrm{Trap}$ references
the intermediate monitor to the trap.  Data for the combined results of these
two steps are plotted as $Q_\mathrm{Telescope}/Q_\mathrm{Trap}$ in
Fig.~\ref{fig:CBPCalib}b.

We therefore measure at each wavelength, three charge ratios:
$Q_\mathrm{CBP}/Q_\mathrm{Telescope}$,
$Q_\mathrm{Telescope}/Q_\mathrm{Monitor}$, and
$Q_\mathrm{Monitor}/Q_\mathrm{Trap}$, each ratio corresponding to one of the
three measurements steps. The product of these gives
$Q_\mathrm{CBP}/Q_\mathrm{Trap}$,

\begin{equation}
\frac{Q_{\mathrm{CBP}}}{Q_{\mathrm{Trap}}} = \frac{Q_{\mathrm{CBP}}}{Q_{\mathrm{Telescope}}}\frac{Q_{\mathrm{Telescope}}}{Q_{\mathrm{Monitor}}}\frac{Q_{\mathrm{Monitor}}}{Q_{\mathrm{Trap}}}  
\end{equation}

The CBP calibration factor $T_\mathrm{CBP}(\lambda)$, defined at each wavelength
as the number of photons out of the CBP per number of photoelectrons measured by
the CBP monitor photodiode, is thus given by:

\begin{equation}
T_{\mathrm{CBP}}(\lambda) = \frac{Q_{\mathrm{Trap}}}{Q_{\mathrm{CBP}}}\frac{1}{EQE_{\mathrm{Trap}}}
\end{equation}

where $EQE_{\mathrm{Trap}}$ is the external quantum efficiency of the trap
detector at each wavelength as calibrated by POWR. The resulting calibration
factor is plotted in Fig.\ref{fig:CBPCalib}c.

Since this paper aims at demonstrating the ability of a CBP to measure the
chromatic variations of telescope and filter transmissions, we don't keep track
of the absolute grey scale but only of $T_\mathrm{CBP}$ up to an arbitrary grey
scale, without propagating the ratios of the photodiodes quantum efficiencies.

The NIST trap responsivity is known with a standard uncertainty of about 0.1 \%.
Statistical standard uncertainty of the charge ratio measurements in each of the
three steps of CBP calibration described above were 0.1 \%, 0.5 \%, and 0.5 \%,
respectively, in the region between 400 nm and 700 nm, and 0.1 \%, 1 \%, and 1
\%, respectively, in the region $\lambda >$ 700 nm.  Uncertainty from systematic
effects such as CBP scattered light, telescope uniformity and scattered light,
spectral purity, and wavelength calibration have not been estimated carefully,
but could add another few percent.  An other potential source of systematic
error is that this procedure uses the 500 $\mu$m pinhole, and thus relies on the
assumption that changing from a large, 500 $\mu$m pinhole in the CBP to a
different pinhole setup with different sizes is achromatic. We list these
effects for completeness but don't add them to the quantitative estimate of
systematics of the later telescope transmission measurements with this CBP. 

We believe the periodic structure seen in the telescope-trap ratio for
$\lambda<400$ nm and $\lambda>700$ nm to be due to interference in the AR
(anti-reflection) coating of the telescope lenses.  This is not seen in the
first step of measurements because the collimated laser beam used in the second
step of measurements is significantly smaller (5 mm) than the exit pupil
diameter of the CBP (approximately 45 mm).  The larger illumination region in
the first step averages the signal over enough different de-phased regions such
that the periodic structure is undetectable.  To reduce the effect of this
systematic on our analysis, we apply a low-pass filter to the data in the
affected regions, with the result shown in red in Fig.~\ref{fig:CBPCalib}c.  The
sharp cutoff for $\lambda<400$ nm likely arises due to the AR coating of the CBP
collimating lens.  As we will show, the uncertainty in the CBP's calibration
becomes the dominant term for wavelengths longer than approximately 800 nm, and
represents an area of significant potential improvement where we expect our new
scheme using solar cells to yield significatively better results.

\section{CBP measurements of the \sd telescope transmission}
\label{sec:CBPExperimentSetup}
\subsection{The \sd experiment}
\label{subsec:SDExperiment}
The \sd experiment aims at anchoring fluxes of currently used spectrophotometric
standard stars (SPSS) to the scale defined by POWR.  In order to reconstruct
fluxes that are free of atmospheric effects, the experiment envisions a long
duration follow-up of SPSS ground level fluxes with a dedicated telescope, whose
absolute full pupil transmission is monitored for the duration by observations
of a calibrated artificial star with NIST traceability. The strategy is to
average out the atmospheric transmission variations, given a long enough lever
arm in time, and to use slitless spectro-photometry to constrain the absorption
spectrum of the atmosphere. The details of this procedure will be left to a set
of forthcoming papers (StarDICE collaboration, 2023-2024), but are expected to
build up on the successed achieved by cosmological supernovae surveys, as for
exampled described in \cite{Burke2017}.

A simple approach for the artificial star is to achieve quasi-perfect plane-wave
illumination of the telescope by the observation of spherical waves emitted by
small light sources in the long (but finite) distance range.  In practice, the
exercise is easier to achieve for small apertures, for which the long range
remains within 200 m.

A test of the proposed concept was performed in 2018 at the Observatoire de
Haute Provence (OHP) with a 10'' f/4 Newtonian telescope. The focal plane of the
telescope was equipped with a SBIG ST-7XME camera behind a 5 slot filter wheel
with $bvRI$ filters and 1 empty slot.  The $b$ and $v$ filters were interference
filters, while the $R$ and $I$ filters were colored glass.  The camera sensor is
a grade 1 front-illuminated Kodak kaf-0402ME CCD, cooled to $\sim -10~^\circ$C
using a Peltier junction for cooling.  The $6.91$~mm$\times4.6$~mm active area
is divided into $765 \times 510$ regions of $9~\mu$m$\times9~\mu$m pixels.

A set of 18 narrow-spectrum LEDs, attached 113.4~m away from the primary mirror
vertex, were observed at the beginning and end of each observation night to
monitor the evolution of the instrument throughput around 18 different
wavelengths, the rest of the night being devoted to the photometric follow-up of
SPSS in 4 $bvRI$ filters.  The test demonstrated a sub-percent precision of the
nightly calibration by the point-like artificial star, while collecting
photometry for a handful of SPSS with $\sim0.01$ mag photon statistics on
individual images.  The test was put to a halt by a sudden and rapid change in
the apparent instrument throughput.  The subsequent analysis of calibration data
demonstrated that the evolution was entirely attributable to a change of the
sensor readout electronic gain, with no apparent change of the detector quantum
efficiency according to the artificial star observations.

We present here a measurement of the monochromatic throughput of this test
instrument that is necessary to interpret the broadband stellar and LEDs
measurements.  The measurement is occuring a posteriori, using the impaired
sensor.  The sensor readout gain appears to have settled at a different but
stable value, so that the most adverse effect is an increased readout noise, to
around 16 e$^-$. We stress for clarity that none of the measurements done at OHP
are discussed here. We only report the procedure done a posteriori at the lab to
characterize the telescope and filters used in the StarDICE demonstration program.

\subsection{Experimental setup}
The CBP system was set up in the StarDICE lab at LPNHE, with the CBP collimator
approximately 1.5 m away from the \sd primary (see Fig.~\ref{fig:CBPPic}).
Images were taken in dark/light pairs -- for ``dark'' images, the \sd camera
shutter was opened, but the laser shutter remained closed.  Subtraction of these
image pairs then corrects for scattered light from contaminating sources.  For
charge measurement, the CBP electrometer was configured to take 10 readings
before and after closing the laser shutter, which permits measurement of
background photodiode charge levels.  During exposures, the monitor spectrograph
was continually read out and saved at a fixed rate, typically 1 Hz or 10 Hz,
depending on total exposure time.  The telescope CCD's exposure time includes an
additional $t_\mathrm{buffer}=10$ seconds relative to the requested exposure
time from the CBP ($t_{exp}$), in order to allow for communication between the
telescope and CBP systems.  Upon completing the exposure, the photodiode and
spectrograph time-series are stored alongside the image.  The CBP scan procedure
can be summarized as

\begin{enumerate}
    \item Open \sd camera shutter, laser shutter remains closed.  Expose for
      time $t_\mathrm{exp} + t_\mathrm{buffer}$. Close camera shutter.
      Used as background subtraction image for following exposure.
    \item Open \sd camera shutter.  Reset electrometer charge and take 10 charge measurements.  Begin taking spectrometer measurements.
    \item Open laser shutter, expose for time $t_\mathrm{exp}$. The exposure
      time is chosen in order to yield enough signal in the camera detector
      without saturating the photodiode. As will be discussed later on, this
      didn't ensure the spectrograph signal to be unsaturated. This issue has been
      taken care of in the new setup. 
    \item Close laser shutter, take final 10 readings on electrometer, stop spectrometer integration.
    \item Close \sd shutter, move to next wavelength, repeat.
\end{enumerate}

In this setup the grid of pinholes is used, in order to test the multiplexing of
the transmission measurement at many different locations on the camera
detector. 

\section{Data reduction}
\label{sec:CBPReduction}
\subsection{Photometry}
\begin{figure}
\begin{center}
\begin{tabular}{c}
\includegraphics[width=0.75\textwidth]{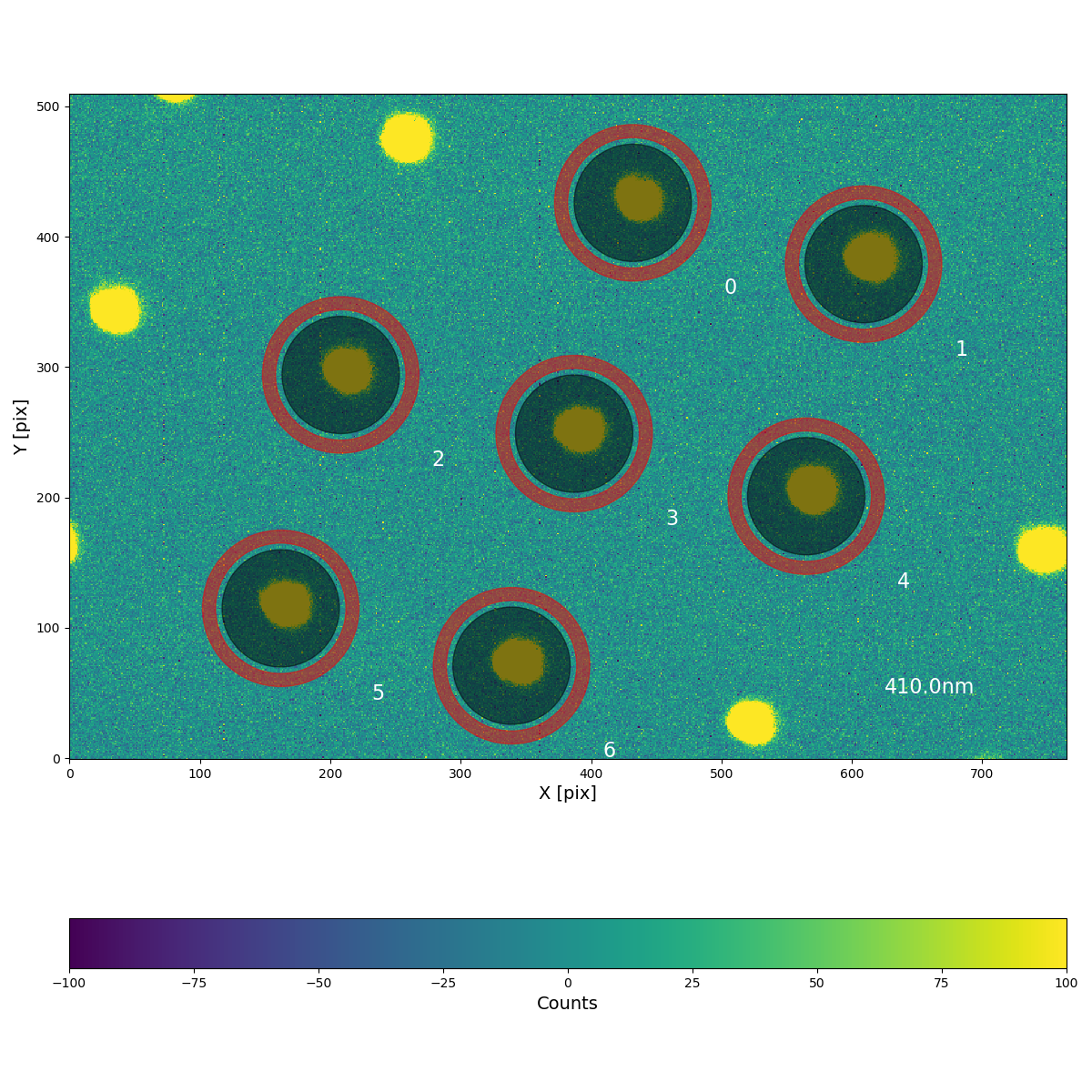}
\end{tabular}
\end{center}
\caption 
{\label{fig:CBPImgEx} An image of the 5x5 20 $\mu$m pinhole mask as seen by the
  \sd CCD.  The image has been bias corrected and differenced with a dark frame.
  Note that a very aggressive stretch is applied in order to see the wings of
  the spots, and that their size in terms of their full-width at half-max (FWHM) is
  much smaller.  The black shaded regions indicate the apertures used for
  photometry, while the red annuli indicate regions used for background subtraction.
  White numbers label each spot, and the wavelength of the laser is shown in the
  lower right.}
\end{figure}

To measure the amount of photons collected by the CCD, we perform aperture
photometry on the grid of points imaged onto the focal plane.  The magnification
ratio between the telescope and the CBP collimating optic is roughly 4; this
means the 20~$\mu$m holes in the CBP mask become 80~$\mu$m diameter spots on the
StarDICE focal plane, which translates to about 9 pixels.  Because there are
refractive optical elements in the telescope beam (e.g., filters, dewar window),
the focus of the system will change slightly with wavelength though the amount
of defocus is not large relative to the size of the spots.  Beyond simple
defocus, during the telescope's trip back to LPNHE from its first calibration
run at OHP (Observatoire de Haute Provence) , we believe that the telescope was knocked slightly out of
collimation, leading to moderate aberrations in the images generated by the CBP.
These aberrations are not prohibitively large, and can be accommodated by
slightly increasing the aperture size used for photometric measurements.
Fig.~\ref{fig:CBPImgEx} shows an example of a bias corrected, dark frame
subtracted image, with the individual aperture and background subtraction
regions marked.  In the analyses that follow, only spots whose entire photometry
and sky background apertures lie entirely within the frame are used.  We use an
aperture of 45 pixels, with a background annulus of inner radius 50 pixels, and a width
of 10 pixels.  When measuring regions of very low transmission, forced aperture
photometry is performed at the last known location of the pinholes.

On occasion, we have noticed changes of a few pixels in the positions of the
pinhole grid across the focal plane.  These arise in particular when changing
filters, though on occasion they seem to arise from presumed slippage in the
alt/az motors driving the CBP pointing.  We compensate for this by allowing the
apertures to move as a fixed grid, with the new locations being determined via
peak finding after convolving the image with a 2-d gaussian with a standard
deviation of 10 pixels in x and y.  Only in the case where all pinholes move
uniformly is the pixel grid allowed to move.  This prevents wandering apertures
in cases where the transmission is effectively zero.

\subsection{Charge measurement}
Photocurrent generated in the monitor photodiode is integrated by the
electrometer, and afterwards is used to normalize away variations in the input
laser light intensity.  We integrate charge rather than measure current in order
to better account for the laser's pulse-to-pulse instability, which can be of
order 10~\% or more.  Measurements are made with the electrometer range fixed to
2 $\mu$C to avoid spurious jumps in charge levels that arise when the
measurement range is changed.  Typical charge levels reached by the CBP during
an integration are (0.1 to 2) $\times 10^{-7}$ C.  Bias current for the Keithley
6514 is specified to be $<4$ fA, which is negligible given our signal levels and
integration times ($\leq300$ s).  Although the specification published in the
6514 datasheet provides only measurement accuracy (not precision), the
manufacturer indicates that the value is intended to be understood as total
measurement uncertainty, including both precision and
accuracy\footnote{Keithley, private communication}.  We therefore use the
datasheet's uncertainty formula for the 2 $\mu$C range given by

\begin{equation}
\label{eqn:KeithleyUncert}
\sigma_\mathrm{PD} = 0.01Q_\mathrm{PD} + 500~\mathrm{pC}
\end{equation}

as the uncertainty on the photodiode charge measurement. We keep the bias part
of the formula in order to account for its variation throughout the experiment
time span. We decided at this level to be conservative and defer the estimate of
the non linearities and stability of the response to the next paper.

\subsection{Spectroscopic Data and Calibration}
The laser's output wavelength is monitored by an Ocean Optics QE65000
spectrograph.  The spectrograph is attached to the CBP integrating sphere by a
600 $\mu$m core diameter optical fiber, which also serves to define the entrance
slit to the spectrograph. We read the device out at a rate of either 10 Hz or 1
Hz, depending on the brightness of the laser at the specified wavelength. The
original choice of the fiber diameter was made in order to maximize the signal
to noise ratio of the spectrograph readings over the full wavelength range.

To calibrate the spectrograph, an Ocean Optics HG-1 wavelength calibration lamp
was shone into the CBP's integrating sphere via an optical fiber, using the same
integrating sphere entrance port as the laser input fiber.  

The spectrograph calibration data has been analysed at the end of our data
taking, after returning the laser to NIST. It was then noticed that the 600
$\mu$m diameter fiber was too broad to properly resolve the lines generated by
the calibration source. Furthermore, we found that there was a systematic,
wavelength-dependent shift in the PSF location between the small and large
fiber, preventing a posteriori recalibration. 

Since it was impossible to redo a full telescope scan with a smaller
spectrograph fiber at that time, we were forced to rely on the stability of
the wavelength provided by the laser given a fixed requested wavelength. Futher
tests on a later CBP implementation with a similar but more recent laser showed
that the relationship between the requested wavelength and the wavelength
provided by the laser is extremely stable. Yet, for this paper we have no
quantitative assessment of a potential shift of the wavelength calibration. 

In summary, the wavelength calibration we use for the current analysis is  as
follows: 
\begin{enumerate}
\item Spectrograph calibration using spectral lamps
\item Calibration of the relation between requested wavelength and the
  wavelength provided by the laser, using the spectrograph with the 100~$\mu$m
  fiber.
\item Use of this relation to transform requested wavelength into actual
  wavelength throughout our analysis of CBP data. 
\end{enumerate}

\subsubsection{Spectrograph Calibration Procedure}
We calibrate the spectrograph by directly linking it to an Ocean Optics Hg-1
lamp with a 100 $\mu$m diameter fiber.

\begin{figure}
\begin{center}
\begin{tabular}{c}
\includegraphics[width=\textwidth]{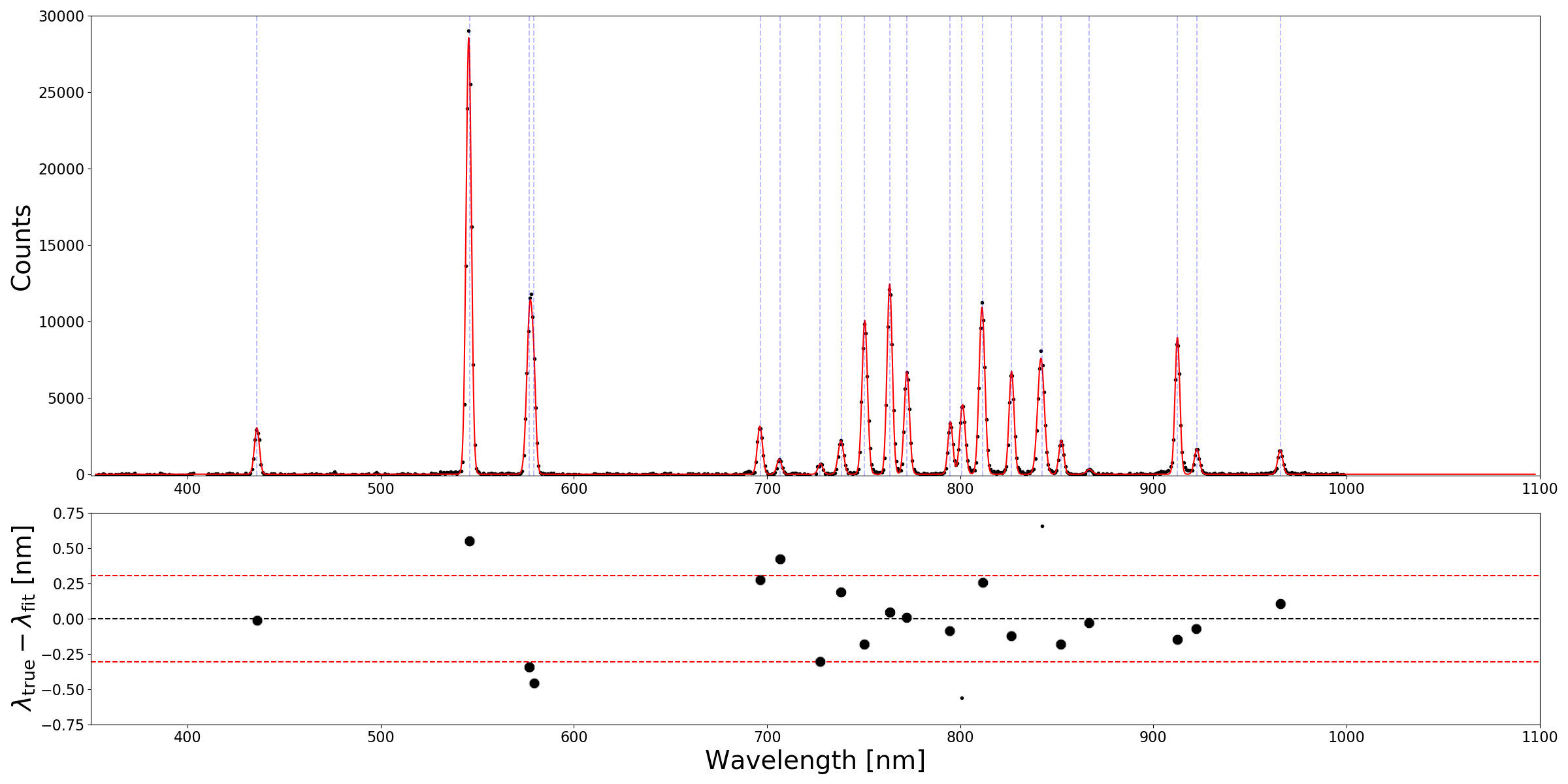}
\end{tabular}
\end{center}
\caption 
{\label{fig:CBPSpecCalib}
Calibration of the CBP monitor spectrograph with an Ocean Optics HG-1 calibration lamp.
On the upper figure, data (black points) is plotted as a function of fitted wavelength and the red curve shows the fitted model.
Pale blue dashed lines denote the true wavelengths of the calibration lines used in the fit.
Data for $\lambda>1000$ nm are masked.
The lower figure shows the difference between true and best-fit wavelengths for the calibration lines.
Red dashed lines are drawn at $+/- 1\sigma$ (= 0.32 nm) from the mean (dashed black line).
}
\end{figure}

We fit the calibration lamp spectra in pixel space using a model that is the sum
of 20 gaussians (one per emission line) and a 0$^\mathrm{th}$ order polynomial
for background estimation, and the results of the fit are shown in
Fig.~\ref{fig:CBPSpecCalib}.  We also note that there is a distinct lack of
calibrated emission lines in the HG-1 lamp for $\lambda\gtrsim950$ nm, thus the
wavelength solution should not be trusted much beyond these values.
Because our CBP calibration data set does not presently extend beyond 1000 nm,
we mask the data for which $\lambda > 1000$ nm (estimated \textit{a priori} from
the spectrograph's built-in wavelength solution).  To determine a
wavelength-pixel mapping for the spectrograph, we fit the centers of the
gaussians against the known calibration lines using a 3$^\mathrm{rd}$ order Chebyshev polynomials.  The per-line residuals of this fit are shown in the lower
panel, and the standard deviation of the lines about 0 indicates that the
wavelength solution is good to approximately 0.3 nm.

\subsubsection{Calibration of the requested wavelength to calibrated wavelength
  relationship for the laser}

\begin{figure}
\begin{center}
\begin{tabular}{c}
\includegraphics[width=\textwidth]{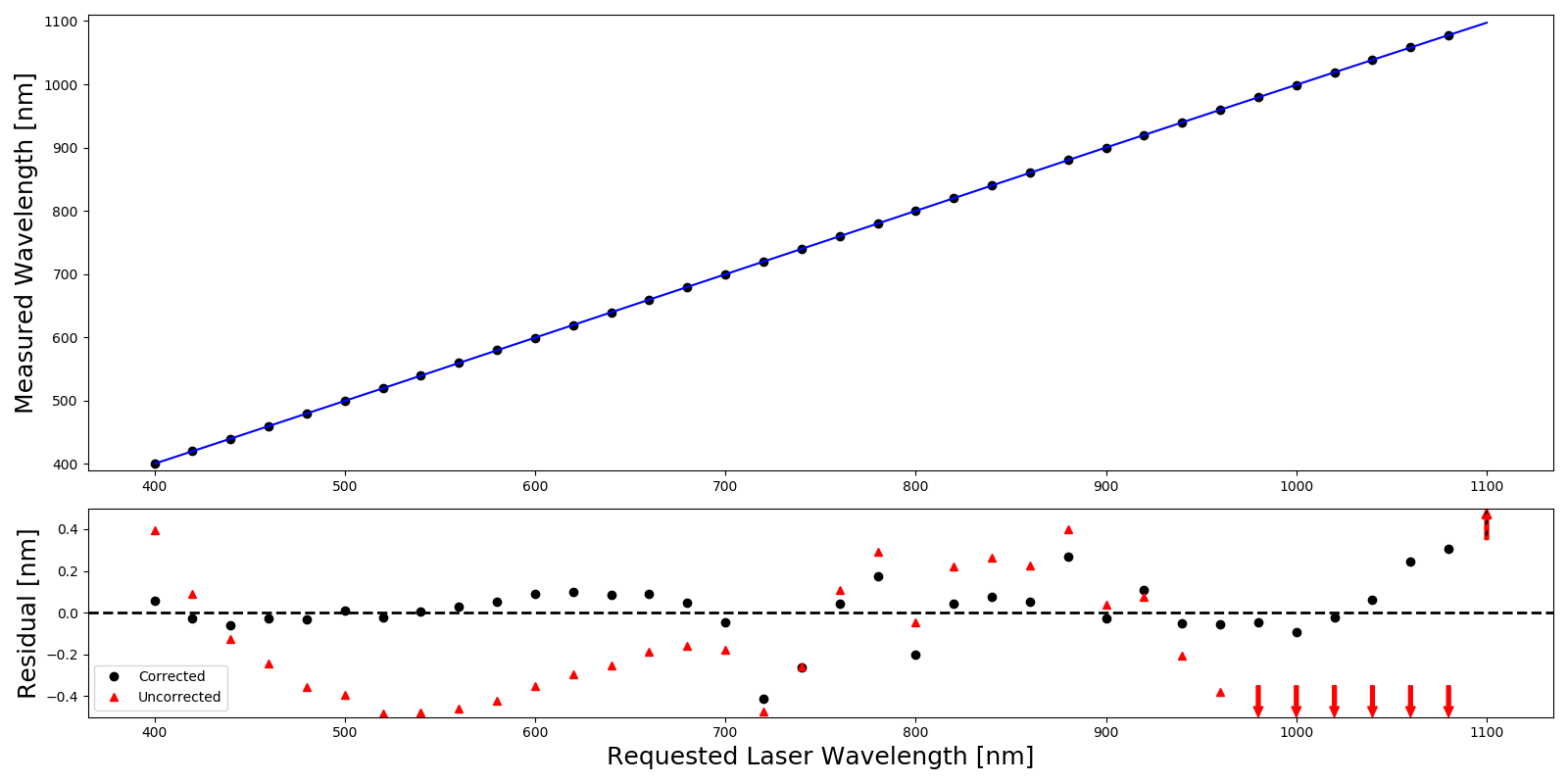}
\end{tabular}
\end{center}
\caption 
{\label{fig:SmallFiberFit} Upper panel: raw data (black dots) and fitted
  Chebyshev polynomial (blue line) showing the spectrograph measured wavelength
  as a function of wavelength requested from the laser.  Lower panel: residuals
  from the fit (black points) as well as the original difference between
  resquested and measured wavelength (red triangles).
  The fit shows residuals at around the 0.1 nm level for most of the wavelength
  range of interest.  Residuals that lie outside the bounds of the lower plot
  are denoted with arrows.  The black and red hatched arrow indicates that the
  residual at 1100 nm lies outside the plot for both the corrected and
  uncorrected case.  }\end{figure}

We transfer this wavelength calibration to the laser by taking a series of laser
output spectra, changing the laser wavelength by 20 nm between each spectra.  We
fit a gaussian to the laser line and then use a 3rd order Chebyshev polynomial
to generate a mapping between requested laser wavelength and spectrograph
measured wavelength.  The results of this fitting process are shown in
Fig.~\ref{fig:SmallFiberFit}.

After applying the polynomial relationship between requested and measured
wavelength, the residuals improve to approximately the 0.1 nm level.  For the
rest of the data presented here we use this polynomial calibration to infer the
true output wavelength of the laser for a given requested wavelength.

We are aware of, and note the fact that the wavelength calibration uncertainty
quoted here is but an estimate, that suffers from the limited amount of time and
data available to write this report. A more detailed propagation of the
wavelength calibration errors will be included in the next version of the
experiment, that will be reported in the next forthcoming paper (Souverin et
al. in prep.).

\subsection{Throughput Calculation}
\label{subsec:ThroughputCalc}
Mathematically, $T_{i}(t, \lambda, \textbf{x}) = Q_\mathrm{CCD,i}(t, \lambda,
\textbf{x})/(Q_\mathrm{PD}(\lambda) T_\mathrm{CBP}(\lambda))$, where $T_{i}$ is
the (unnormalized) telescope transmission as seen by pinhole $i$,
$Q_{\mathrm{CCD},i}$ is the charge collected by the CCD for pinhole $i$,
$Q_\mathrm{PD}(\lambda)$ is the charge collected by the CBP monitor photodiode,
and $T_\mathrm{CBP}(\lambda)$ is the transmission correction for the CBP system
as outlined in Section~\ref{sec:CBPCalib}.  Because the true quantum efficiency
of the CBP monitor photodiode is rolled into our CBP calibration term, there is
no correction factor for the CBP monitor photodiode.  The different
pinholes allow for a degree of focal plane multiplexing by sampling different
sections of the system within a single exposure. This provides potentially the
ability to measure grey and chromatic variations of the full detector response. 

To show the derivation for this form of $T_i$, we can first use
Eqn.~\ref{eqn:CBPsignal} to write
\begin{equation}
Q_{\mathrm{CCD},i}=\sum_{\mathrm{pix},i}\phi_\mathrm{CBP}(t, \textbf{x},
\lambda)\Delta\,t
\end{equation}
where the sum is over the pixels belonging to the image of
pinhole $i$, which serves to localize the measurement to the region around point
$\textbf{x}$ on the focal plane, and $\Delta t$ is the integration time.  We
then arrive at the expression

\begin{equation}
    Q_{\mathrm{CCD},i}(t, \lambda, \textbf{x})= \sum_{\mathrm{pixels},i} C(t) T(t, \lambda, \textbf{x}) S_\mathrm{pixel}(t, \lambda, \textbf{x}) \Delta\,t
\end{equation}

where $S_\mathrm{pixel}$ is the photon flux seen by a given pixel in aperture
$i$ such that $\sum_\mathrm{pixels} S_\mathrm{pixel}(t,\lambda, \textbf{x}) =
S(t, \lambda,)$, and we have invoked the approximation that the CBP is sampling
a representative weighting of the \sd primary (allowing us to drop the
$\textbf{r}$ dependence), enabled by the coincidence that the CBP output beam
size is roughly the size of the \sd primary annulus (We define the primary
annulus as the primary disc hollowed out of the shade of the secondary).  The
calibration factor of the CBP monitor photodiode $T_\mathrm{CBP}$ has been
transformed up to an arbitrary grey term, that we set to one for simplicity,
asserting that each photoelectron corresponds to a single unique detected
photon.  The total number of photons entering the \sd telescope for a given
pinhole, written as $S(\lambda, t) \Delta\,t\,K_\mathrm{i}$, is also given by
$Q_\mathrm{PD} T_\mathrm{CBP}$, where $K_i$ is an unknown constant reflecting
our ignorance of the true size of the emitting pinhole.  By assuming that the
transmission of the telescope system is reasonably flat over the size of the
pinhole image on the detector, we can write the ratio
$Q_\mathrm{CCD,i}/(Q_\mathrm{PD} T_\mathrm{CBP})$ as

\begin{equation}
    \frac{Q_\mathrm{CCD,i}}{Q_\mathrm{PD} T_\mathrm{CBP}} = \frac{C(t)T(t,\lambda,\textbf{x}_\mathrm{center}) \sum_\mathrm{pixels,i} S_\mathrm{pixel}(t,\lambda) \Delta\,t}{K_\mathrm{i} S(t, \lambda) \Delta\,t} = \frac{C(t)}{K_\mathrm{i}} T(t, \lambda, \textbf{x}_\mathrm{center})
\end{equation}

where the term $\frac{C(t)}{K_\mathrm{i}}$ is a constant with respect to
wavelength.

Under the assumption that the temporal variability of $C(t)$ (which is mostly
related to variables such as electronic gain and grey extinction due to e.g.,
dust on the primary) is slow compared to the time needed to measure the
transmission, we can treat the ratio as a constant, $C$.  This constant is the
same for all measurements of a given pinhole $i$ at location $\textbf{x}$ for a
given scan, and we can normalize it away by dividing by the transmission at a
fiducial wavelength or by the average transmission over all wavelengths.  This
gives us a measurement of system transmission at time $t$ and focal plane
position $\textbf{x}$ relative to another wavelength, thus reaching our goal of
measuring the relative throughput of the system.

In summary, the process of measuring throughputs with a CBP consists of
performing aperture photometry on each of the pinhole images, dividing by the
photodiode charge measurement and CBP calibration factor at that wavelength, and
then normalizing by, for example, the average of that quantity over all
wavelengths. We then obtain a set of relative transmissions, one per pinhole $i$,
covering the entire detector, allowing to trace both the average transmission
for a given CBP partial illumination of the primary mirror, and potential
chromatic transmission variations over the field of view. 

\subsection{Throughput uncertainty calculation}
\label{subsec:UncertCalc}
\begin{figure}
\begin{center}
\begin{tabular}{c}
\includegraphics[width=0.75\textwidth]{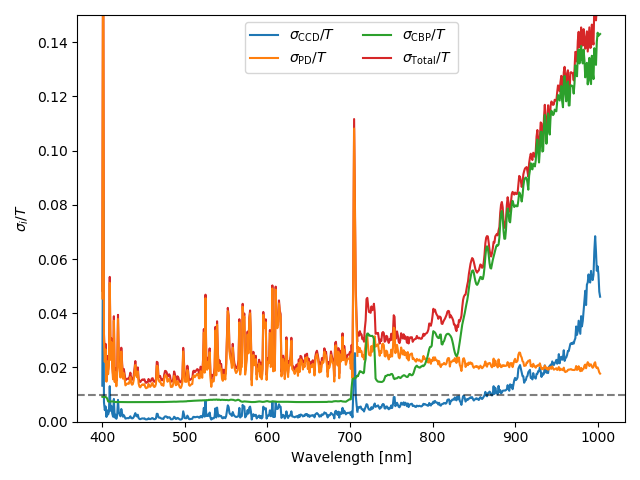}
\end{tabular}
\end{center}
\caption 
{\label{fig:CBPUncertBudget} The uncertainty budget for the CBP measurement of
  \sd throughput.  Contributions to the total uncertainty budget (red) are shown
  for the 3 main components: aperture photometry (blue), photodiode charge
  measurement (orange), and CBP calibration (green).  The spike seen near 700 nm
  is due to the degeneracy region in the laser, and the spike near 400 nm is due
  to low laser brightness combined with low CBP optical throughput.  The gradual
  rise from approx. 850 nm onward is believed to be due to the interference
  fringing effect discussed in Sec.~\ref{sec:CBPCalib}.  }
\end{figure}

Using the standard method for uncertainty propagation, the uncertainty on the
throughput measurement $T = Q_\mathrm{CCD}/(Q_\mathrm{PD}T_\mathrm{CBP})$ is
given by

\begin{equation}
\label{eqn:TotalUncert}
    \sigma_T = \big( (\frac{\partial T}{\partial Q_\mathrm{CCD}} \sigma_\mathrm{CCD})^2 + (\frac{\partial T}{\partial Q_\mathrm{PD}} \sigma_\mathrm{PD})^2 + (\frac{\partial T}{\partial T_\mathrm{CBP}} \sigma_\mathrm{CBP})^2 \big)^\frac{1}{2}
\end{equation}

where $\partial T/\partial Q_\mathrm{CCD} = 1/(Q_\mathrm{PD} T_\mathrm{CBP})$,
$\partial T/\partial Q_\mathrm{PD} = -Q_\mathrm{CCD}/ (Q_\mathrm{PD}^2
T_\mathrm{CBP})$, and $\partial T/\partial T_\mathrm{CBP} =
-Q_\mathrm{CCD}/(Q_\mathrm{PD} T_\mathrm{CBP}^2)$.  The uncertainty budget for
the CBP's measurement of the \sd telescope's no-filter throughput is shown in
Fig.~\ref{fig:CBPUncertBudget}.  For wavelengths below approximately 800 nm, the
uncertainty is dominated by uncertainty in the charge measurement.  From the
definitions given above (equations~\ref{eqn:KeithleyUncert}
and~\ref{eqn:TotalUncert}), it is easy to show that the minimum uncertainty
contribution from the photodiode term is 1\%, in the limit that $50 \mathrm{nC}
<< Q_\mathrm{PD}$.  Once the prescribed limit is reached, the photodiode
uncertainty is not a function of total charge, and therefore decouples from our
ability to reduce it by exposing longer, or using a brighter source.  It can be
reduced only by making more measurements.  At wavelengths longer than $\sim790$
nm, the CBP's flux calibration (Section~\ref{sec:CBPCalib}) is the limiting
factor. In particular, the uncertainty ascribed to the interference fringing in
the calibration transfer telescope.

\section{Results and Discussion}
\label{sec:CBPResults}

\subsection{\sd telescope transmission}
\begin{figure}
\begin{center}
\begin{tabular}{c}
\includegraphics[width=\textwidth]{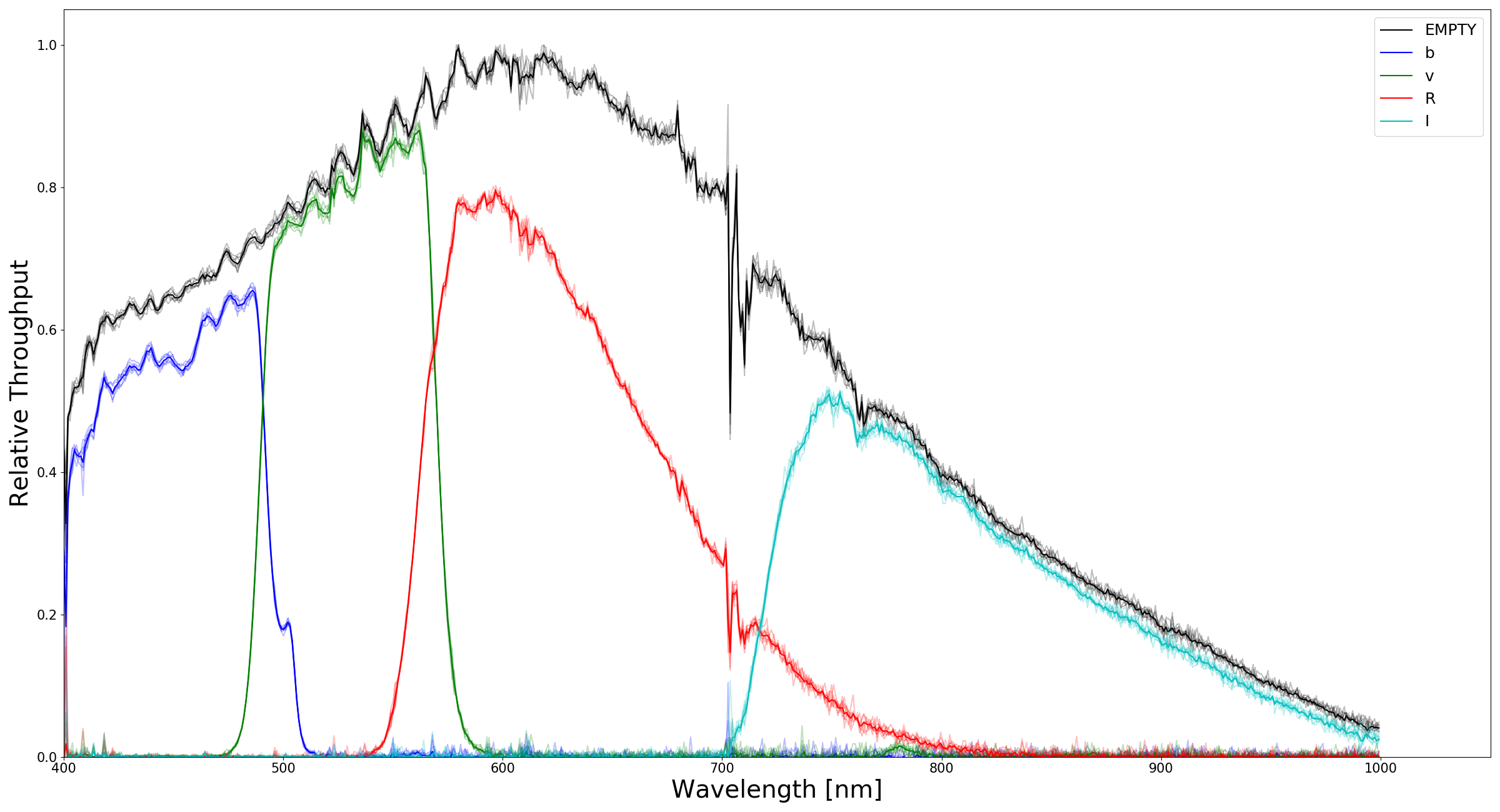}
\end{tabular}
\end{center}
\caption 
{\label{fig:SDThroughput} A multi-pinhole CBP scan of the \sd telescope.  Each
  color represents a different filter placed in the beam of the telescope.  The
  transparent curves show the measured transmissions for each of the pinholes in
  the grid, while the solid curves are the means across all pinholes.  Each
  pinhole is normalized relative to its own peak value in the no-filter scan.
  The sharp features seen around 710 nm are due to extremely low SNR
  (signal-to-noise ratio), on account of the degeneracy region of the laser.}
\end{figure}

Fig.~\ref{fig:SDThroughput} shows the transmission of the \sd telescope as
measured by the average over all pinholes within a single CBP pointing.  In
order to examine the relative throughput of the system, we normalize each
pinhole by its peak value with no filter in the beam, thereby allowing us to
directly compare transmission variations across pinholes.  The periodic
variations seen in the transmission curves in the blue ($\lambda \lesssim 600$
nm) are believed to be real, and due to interference fringing in the microlens
array mounted on the \sd CCD.  Although not visible on the results plotted, we
mention for completeness that there is also one point (810 nm, $I$ band) that is
excised due to laser issues during the exposure.

The data ends at 1 $\mu$m, as this is the cutoff wavelength for the trap
detector calibration we currently have in hand.  As the goal of this paper is to
outline the performance of the CBP, we do not extrapolate the curve farther, nor
implement an interpolation scheme for the degeneracy region and instead leave
these elements of the StarDICE bandpass analysis to the forthcoming \sd
experiment paper (StarDICE collaboration 2023-2024).  In general the filters are
well-behaved with one exception 
being a small leak in the $v$ filter around 790~nm.  The sharp features around
710~nm are due to the degeneracy region of the laser, and are not true features
of the \sd bandpass.

\subsection{Evidence for variation across StarDICE focal plane}
\begin{figure}
\begin{center}
\begin{tabular}{c}
\includegraphics[width=\textwidth]{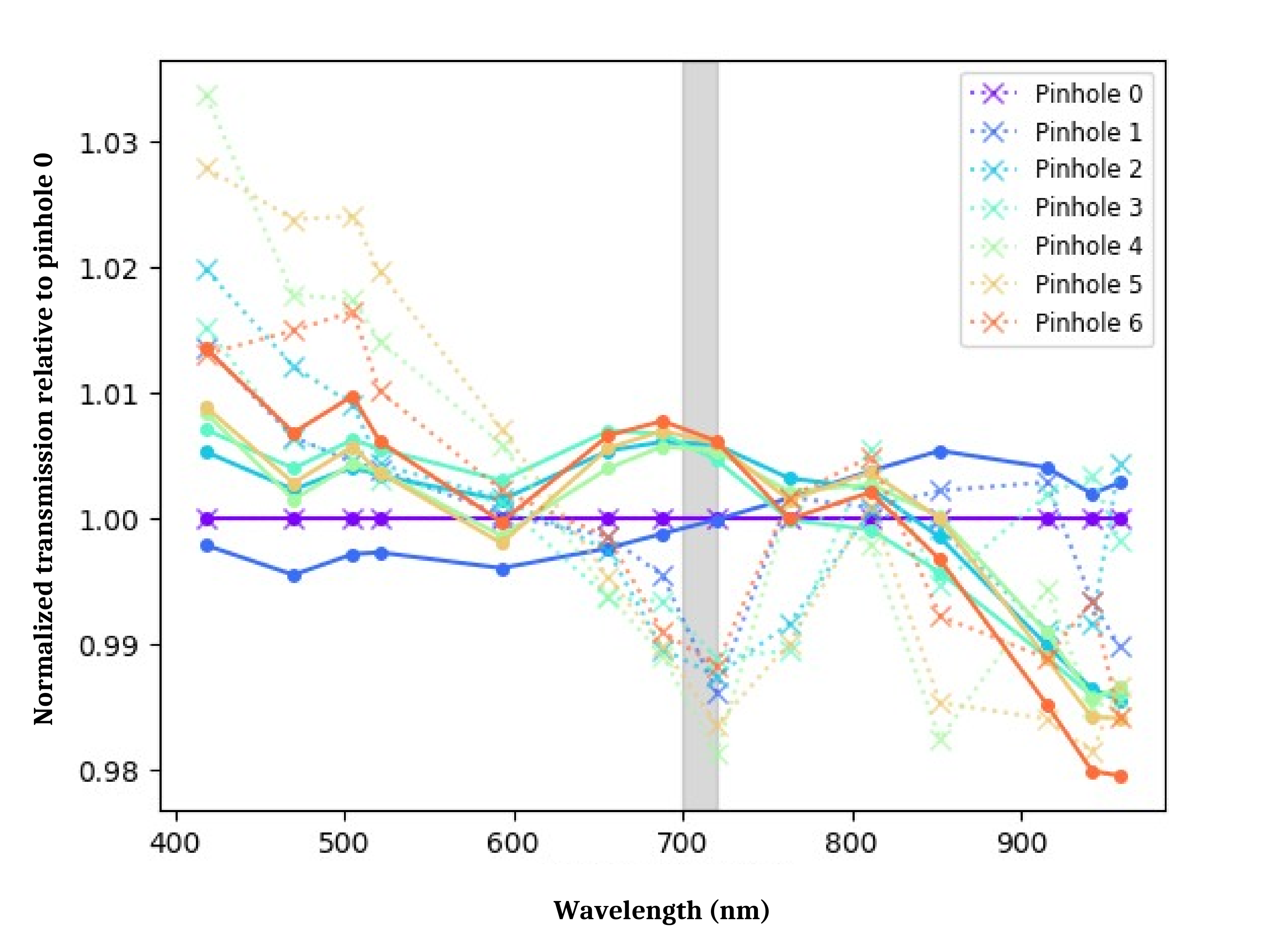}
\end{tabular}
\end{center}
\caption 
{\label{fig:SDSpatialThroughput} Normalized transmissions for each of the
  pinholes from the CBP scan of the \sd telescope.  Xs denote CBP measurements
  integrated over the wavelength width of the LEDs used in the flat field study
  of the CCD.  Circles signify LED flat field illumination measurements of the
  camera, integrated over 10 x 10 pixel squares at the position of the CBP
  pinholes.  The grey strip denotes the degeneracy region, where CBP
  measurements are less reliable.  Some evidence for spatial variability in the
  transmission function is seen, particularly in the blue end of the spectrum.}
\end{figure}

In addition to the average transmission measurement of
Fig.~\ref{fig:SDThroughput}, we report a chromatic variation of roughly 5~\% for
the CBP open transmission measurement (no filter) on a per-pinhole basis.  This
is shown on Fig.~\ref{fig:SDSpatialThroughput}, where the crosses represent CBP
measurements of the StarDICE telescope transmission over the full optical range,
colored by pinhole number.  They have been integrated over narrow ($\sim 30$ nm)
bands for later comparison with LED flat field measurements of the camera
transmission.  We see that, relative to the measurement for pinhole 0, every
other pinhole displays a chromatic evolution of the open transmission of about
5~\% from 400~nm to 950~nm.

We also observe a systematic difference between pinholes, with pinholes 4-6
being more transmissive than others at 400~nm and less transmissive at 950~nm.
This corresponds to a spatial evolution over the field of view of about 5~\%.

Part of these effects can be accounted for by the camera response, as shown by
the round coloured dots on the same figure.  Those points represent the open
transmission of the camera measured by flat field illumination using LEDs of
narrow ($\sim 30$~nm) spectra.  The flat fields obtained are integrated over
spatial regions corresponding to each pinhole and displayed in the same color
coding as the CBP measurements.  We see that both the chromatic trend of all
pinholes and the spatial trend between pinholes are partially accounted for by
the response of the camera.  A more quantitative assessment of the spatial
variation detected would need a full range of new measurements, both with the
CBP and with the flat field illumination system, which is beyond the scope of
this paper. It nonetheless shows that a CBP is a promising tool to measure the
response of the full telescope system down to the camera, even at the level of
variations within the field of view. 

\subsection{Reproducibility}
\begin{landscape}
\begin{figure}
\begin{center}
\begin{tabular}{c}
\includegraphics[angle=-90, origin=c, height=\textwidth]{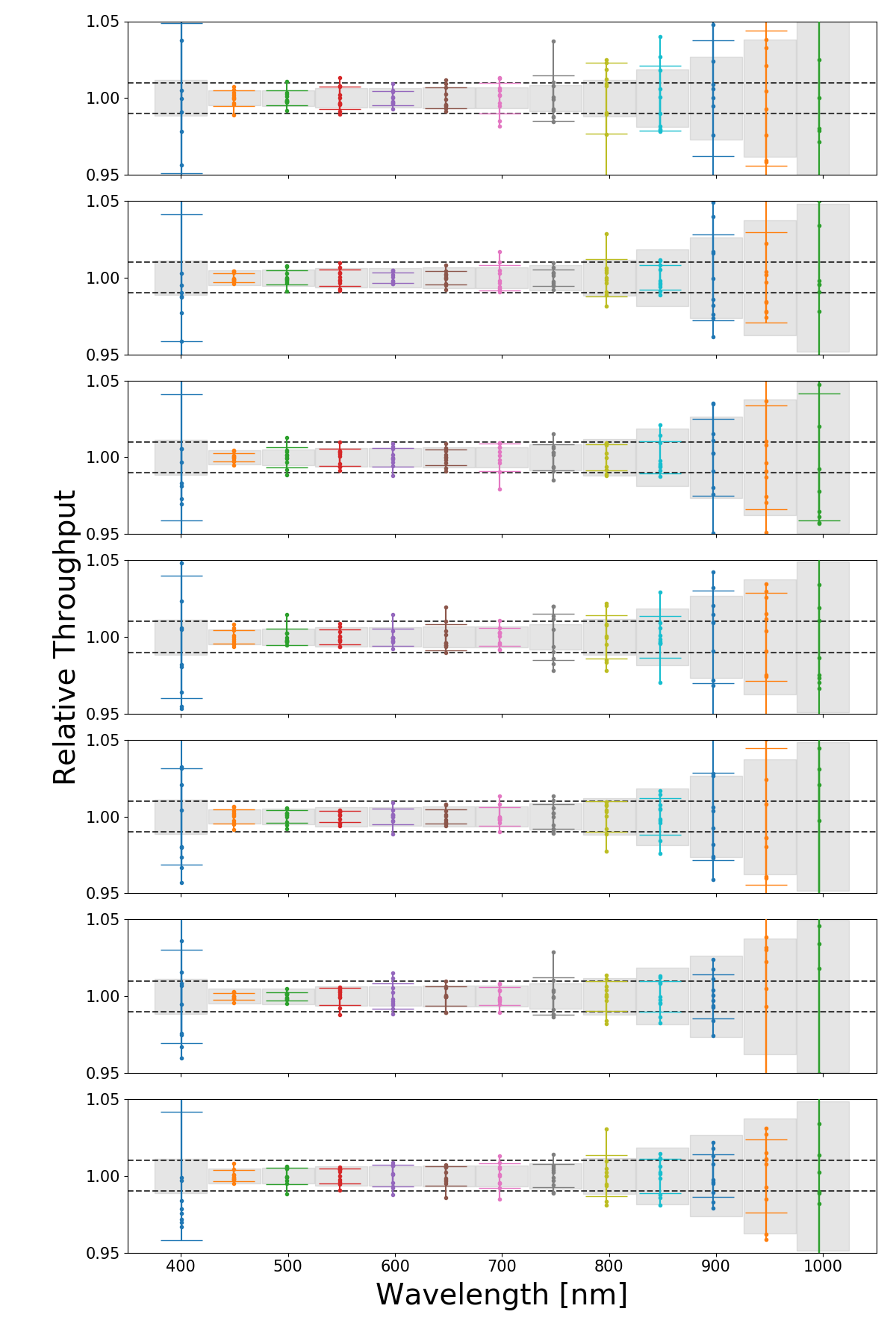}
\end{tabular}
\end{center}
\caption 
{\label{fig:CBPRepeatability} Repeatability test of the telescope transmission
  measurement from 400 nm - 1000 nm at 50 nm intervals.  Each plot represents an
  individual pinhole on the \sd focal plane.  The horizontal colored bars at
  each wavelength represent +/-1$\sigma$ from the mean of the measurements.
  Each data point is normalized by the mean of the measurements at that
  wavelength for that pinhole spot.  For some of the measurements, the
  +/-1$\sigma$ lines fall slightly outside the limits of the graph.  The dashed
  grey lines denote +/-1 \% deviation from the mean.  Solid grey regions denote
  predicted uncertainty using the total uncertainty budget from
  Fig.~\ref{fig:CBPUncertBudget}.  }
\end{figure} 
\end{landscape}

In addition to the uncertainty calculations given in
Section~\ref{subsec:UncertCalc}, we also obtained repeated measurements of the
transmission function of the telescope at a fixed subset of wavelengths to test
the robustness of the CBP results.  Fig.~\ref{fig:CBPRepeatability} shows the
results of 10 measurements of the \sd system throughput taken at 13 different
wavelengths (50~nm intervals from 400~nm to 1000~nm) on a per-pinhole basis.
Note that the pinholes and focal plane locations are not the same as those in
Fig.~\ref{fig:SDThroughput}.  The measurements were taken as 10 sets of 13
measurements (i.e., the wavelength was changed between each measurement), which
temporally separates the measurements at each wavelength by several minutes.
Each pinhole-wavelength combination is normalized to its mean value, and
horizontal colored lines are drawn at $+/-1\sigma$.  Black horizontal lines
denote the calibration uncertainty goal of 1~\% (standard uncertainty), which is
generally met between approximately 450~nm to 850~nm.  This is in line with the
theoretical predictions, which say that uncertainty should decrease by
$1/\sqrt{N}$.  For $N=10$ measurements, we would expect 1~\% precision for
measurements with individual uncertainties of $\sim3~\%$ -- this corresponds
roughly to the 450~nm to 800~nm region in Fig.~\ref{fig:CBPUncertBudget}. This is
because the uncertainty of the CBP photodiode charge measurements is statistics
dominated in that spectral region. Where the systematic error due to the CBP
transmission uncertainty dominates, the uncertainty doesn't decrease with
repeated measurements.

The grey shaded regions are visualizations of this uncertainty for each
wavelength.  The under-estimation of the predicted uncertainty and the observed
scatter is likely due to the very sharp nature of the AR coating cutoff around
400~ nm, and the calibration consequences thereof.

\subsection{Additional Sources of Systematic Error}
We have attempted, where practical, to maintain an accurate uncertainty budget
for the CBP system.  Here we discuss other possible sources of systematic
uncertainty which might affect our measurements, and offer some possible future
avenues of estimation or suppression.

One assumption we have made in the design of the CBP is the achromaticity of the
pinhole grid illumination by the integrating sphere.  If the back illumination
of the grid is chromatic across the grid, that would directly translate to
spurious chromatic evolution across the detector on the \sd system.  Examining
this would involve rotating or translating the detector under examination with
respect to the CBP grid.  By comparing data taken with the same spot at
different areas of the focal plane (or different spots at the same area of the
focal plane), we can place constraints on the achromaticity of the CBP pinhole
illumination.  Alternatively, if one wishes to quash this source of uncertainty
at the expense of multiplexing, a single on-axis pinhole can be used, which will
not suffer from this effect.  We have plans to continue to explore this
potential systematic on the \sd system in the future.

Another assumption is that the CBP calibration factor, which was measured at
NIST with a single on-axis 500 $\mu$m pinhole (for SNR reasons), is directly
transferable to the pinhole grid, with pinholes of significantly smaller sized
and with most (if not all) of them at least slightly off axis.  Measuring
transfer achromaticity directly is challenging because the 500 $\mu$m pinhole
has an area, and therefore signal, that is 25x larger than the total pinhole
area of the 5x5 20 $\mu$m pinhole grid (and 625x larger than any individual
pinhole).  Repeating the NIST calibration process with this much less flux is a
challenging proposition, and out of the scope of this paper. The assumption of
calibration invariability with pinhole size will be a source of systematics
investigated in details in the forthcoming paper (Souverin et
al. in prep.).

Electronic gain variation represents another axis of systematic uncertainty.
Gain drifts might occur in either the CCD readout electronics or in the
electrometer; in either case, if the gain varies systematically over the course
of a CBP scan, it would show up as a change in system response with wavelength.
There is a method to counter this, which is to consistently return to a
reference configuration (in both pointing and wavelength) during the scan.  By
taking data in the exact same configuration, any temporal drift in the
electronics during the scan can be monitored and regressed away.

\section{Conclusion}
\label{sec:Conclusion}
We have presented here the Collimated Beam Projector (CBP) which is a telescope
transmission calibration device. The CBP uses collimated light to illuminate a
telescope, mimicking a stellar wavefront over a subsection of the primary. By
calibrating the output of the CBP relative to a NIST trap detector, the CBP can
propagate the state-of-the-art POWR calibration to a telescope detector.

To demonstrate this, we have measured the throughput of the \sd telescope, with
uncertainties between 3 \% ($400$ nm $\lesssim\lambda \lesssim 800$ nm) and 14
\% ($800$ nm $\lesssim \lambda \lesssim 1000$ nm). Over the full wavelength
range, the wavelength calibration achieved, relying on the stability of the
laser, is estimated to be of the order of 0.1~nm.

In its present form, precision for the CBP is limited by different factors in
each of the regimes mentioned.  In the short wavelength regime, the accuracy of
the electrometer used to measure charge deposited on the monitor photodiode
limits single-exposure uncertainties to 1 \%, even at high flux.  For the long
wavelength region, we are limited by systematics in the calibration of the CBP,
which arise due to interference fringing in the transfer telescope used in the
calibration process.

To address the former limitation, two possibilities exist: make $N$ measurements
and average, or procure a different, higher precision, electrometer.  As an
example, the Keithley 6517B electrometer has a per-measurement precision of 0.4
\% + 50 pC for the 2 $\mu$C measurement range, which is a significant
improvement over the 6514 used in this work.  Making $N$ measurements to improve
SNR is one of the improvements that have been implemented with success in futher
work. 

In the long wavelength regime, we are considering two avenues of approach:
recalibration at NIST using a different optical setup using a reflective instead
of refractive telescope to avoid fringing issues, or re-designing the
collimation scheme for the CBP such that a monitor photodiode can be inserted
directly to the output beam (rather than the integrating sphere).  This latter
scheme also has the advantage in that it requires only the calibration of the
monitor photodiode relative to POWR, and not the calibration of the full CBP
instrument, and is the path we have selected for our further developments. The
main issue of having an output flux largely diluted at the output of the CBP,
resulting on a signal too low to be accurately measured has been overcome by
the use of a large surface solar cell that collects the entire output beam. The
calibration of such a solar cell is described in \cite{Brownsberger2021} and
its use will be discussed in a forthcoming paper (Souverin et
al. in prep.). 

In general, this implementation of a CBP in front of a telescope in a controled
laboratory setting has provided many useful venues for improvement. Those have
been implemented and will be presented in the forthcoming paper (Souverin et
al. in prep.).

A calibration system that can be taken to different telescopes is of critical
importance to supernova cosmology in particular, as photometric calibration has
been and remains a major factor in determining cosmological
parameters\cite{Scolnic2014, Stubbs2015, Scolnic2018, Betoule2014}.  In support
of this mission, our group intends to undertake a series of measurements on a
representative group of telescopes that have played significant roles in
supernova measurements.  In addition to our CBP, Rubin Observatory has also
procured a CBP.  Measurements and techniques learned here will be directly
translated to support observations taken by the Rubin Observatory, and
ultimately support the transfer of the NIST optical flux scale to astrophysical
objects.

\acknowledgments

This paper has undergone internal review in the LSST Dark Energy Science
Collaboration. We thank P.Antilogus, J.Neveu and N.Regnault for their thorough
job.

The authors declare no conflicts of interest.

C.W.Stubbs performed the initial conceptual design of the CBP, and was engaged
in the implementation and data analysis. M.W.Coughlin designed and fabricated
the original version of the CBP, as well as supported measurements and paper
writing. N.Mondrik conducted the installation of the CBP in the lab, led the
data taking and analysis and wrote the main part of the paper. M.Betoule
contributed to the StarDICE telescope control system and data acquisition
software, and took part in the hardware assembly, data taking and
analysis. S.Bongard participated in the hardware assembly, data taking and
analysis, and took charge of the editorial needs of the paper.  P.S. Shaw and
J.T. Woodward assembled the equipment and NIST and performed the measurements
and data analysis. J.P. Rice wrote the NIST calibration section and provided
editorial support.

The DESC acknowledges ongoing support from the Institut National de 
Physique Nucl\'eaire et de Physique des Particules in France; the 
Science \& Technology Facilities Council in the United Kingdom; and the
Department of Energy, the National Science Foundation, and the LSST 
Corporation in the United States.  DESC uses resources of the IN2P3 
Computing Center (CC-IN2P3--Lyon/Villeurbanne - France) funded by the 
Centre National de la Recherche Scientifique; the National Energy 
Research Scientific Computing Center, a DOE Office of Science User 
Facility supported by the Office of Science of the U.S.\ Department of
Energy under Contract No.\ DE-AC02-05CH11231; STFC DiRAC HPC Facilities, 
funded by UK BEIS National E-infrastructure capital grants; and the UK 
particle physics grid, supported by the GridPP Collaboration.  This 
work was performed in part under DOE Contract DE-AC02-76SF00515.

We thank the US Department of Energy and the Gordon and Betty Moore Foundation
for their support of our LSST precision calibration efforts, under DOE grant
DE-SC0007881 and award GBMF7432 respectively.  NM is supported by the National
Science Foundation Graduate Research Fellowship Program under Grant
No. DGE1745303.  
MC acknowledges support from the National Science Foundation with grant numbers PHY-2010970 and OAC-2117997.
Any opinions, findings, and conclusions or recommendations
expressed in this material are those of the authors and do not necessarily
reflect the views of the National Science Foundation.  Identification of
commercial equipment to specify adequately an experimental problem does not
imply recommendation or endorsement by the NIST, nor does it imply that the
equipment identified is necessarily the best available for the purpose.  This
research made use of Astropy,\footnote{http://www.astropy.org} a
community-developed core Python package for astronomy \cite{Robitaille2013,
  PriceWhelan2018}.

The data used for this paper can be retrieved by direct inquiry to the
corresponding author. Given the demonstrating nature of the work undertaken in
this paper, we do not plan more extensive online publication of the data.


\bibliography{mendeley_bib,report}   
\bibliographystyle{spiejour}   


\vspace{1ex}
\noindent Biographies and photographs of the other authors are not available.

\listoffigures
\listoftables

\end{spacing}
\end{document}